\documentclass[iop,apj,12pt,twocolappendix,tighten]{emulateapj}

\usepackage{acronym}
\usepackage{booktabs}
\usepackage{bm}
\usepackage{dcolumn}
\usepackage[plainpages=false, colorlinks=true, anchorcolor=blue, linkcolor=blue, citecolor=blue, bookmarks=false]{hyperref}
\usepackage[utf8]{inputenc}
\usepackage{lipsum}
\usepackage[super]{nth}
\usepackage{soul}
\usepackage{xspace}

\usepackage{natbib}
\newcommand{\myhead}[1]{\multicolumn{1}{c}{#1}}
\newcolumntype{d}[1]{D{.}{.}{#1}}

\newcommand{\fornax}{\textsc{Fornax }}

\newcommand{\dd}{\mathrm{d}}

\newcommand{\ie}{\textit{i.e.}}
\newcommand{\eg}{\textit{e.g.}}

\begin{document}
\slugcomment{Draft version \today}

\title{Electron-Capture and Low-Mass Iron-Core-Collapse Supernovae:\\
New Neutrino-Radiation-Hydrodynamics Simulations}
\author{David Radice\altaffilmark{1,2}, Adam Burrows\altaffilmark{2},
David Vartanyan\altaffilmark{2}, M. Aaron Skinner\altaffilmark{3}, and
Joshua C. Dolence\altaffilmark{4}}
\altaffiltext{1}{Institute for Advanced Study, 1 Einstein Drive, Princeton,
NJ 08540, USA}
\altaffiltext{2}{Department of Astrophysical Sciences, Princeton
University, 4 Ivy Lane, Princeton, NJ 08544, USA}
\altaffiltext{3}{Lawrence Livermore National Laboratory, 7000 East Ave.,
Livermore, CA 94550-9234}
\altaffiltext{4}{CCS-2, Los Alamos National Laboratory, P.O. Box 1663
Los Alamos, NM 87545}

\begin{abstract}
We present new 1D (spherical) and 2D (axisymmetric) simulations of
electron-capture (EC) and low-mass iron-core-collapse supernovae (SN).
We consider six progenitor models: the ECSN progenitor from
\citet{nomoto:1984a, nomoto:1987a}; two ECSN-like low-mass
low-metallicity iron core progenitors from Heger (private
communication); and the 9-, 10-, and $11$-$M_\odot$ (zero-age main
sequence) progenitors from \citet{sukhbold:2016a}.  We confirm that the
ECSN and ESCN-like progenitors explode easily even in 1D with explosion
energies of up to a $0.15$ Bethes ($1\ {\rm B} \equiv 10^{51}\ {\rm
erg}$), and are a viable mechanism for the production of very low-mass
neutron stars. However, the 9-, 10-,  and $11$-$M_\odot$ progenitors do
not explode in 1D and are not even necessarily easier to explode than
higher-mass progenitor stars in 2D. We study the effect of perturbations
and of changes to the microphysics and we find that relatively small
changes can result in qualitatively different outcomes, even in 1D, for
models sufficiently close to the explosion threshold. Finally, we
revisit the impact of convection below the protoneutron star (PNS)
surface. We analyze, 1D and 2D evolutions of PNSs subject to the same
boundary conditions. We find that the impact of PNS convection has been
underestimated in previous studies and could result in an increase of
the neutrino luminosity by up to factors of two.
\end{abstract}

\section{Introduction}
\label{sec:introduction}
The formation of a massive iron core at the end of the evolution of
stars with \ac{ZAMS} masses larger than ${\sim} 12\ M_\odot$ is a robust
prediction of stellar evolution theory. These stars undergo
core-collapse once their cores reach the Chandrasekhar mass and may
explode as \acp{CCSN}. \acused{SN}

\begin{figure*}
  \begin{minipage}{\columnwidth}
    \includegraphics[width=\columnwidth]{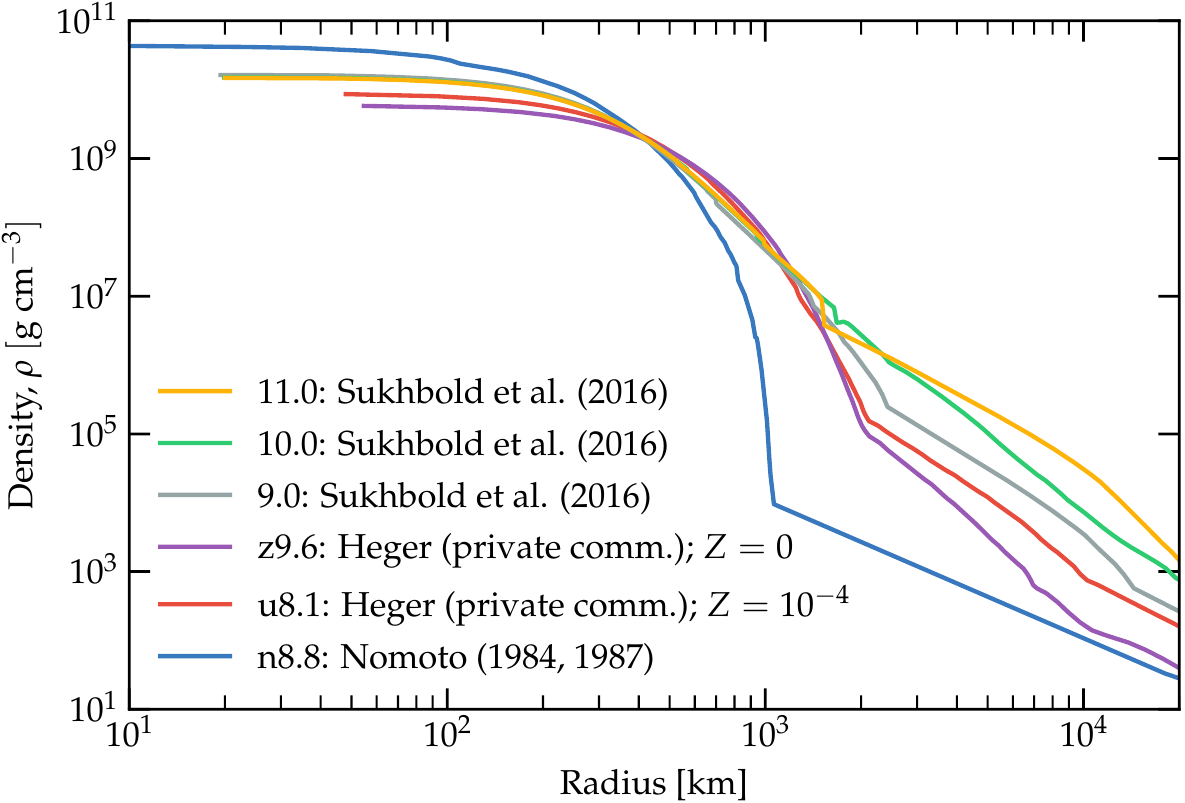}
  \end{minipage}
  \hfill
  \begin{minipage}{\columnwidth}
    \includegraphics[width=\columnwidth]{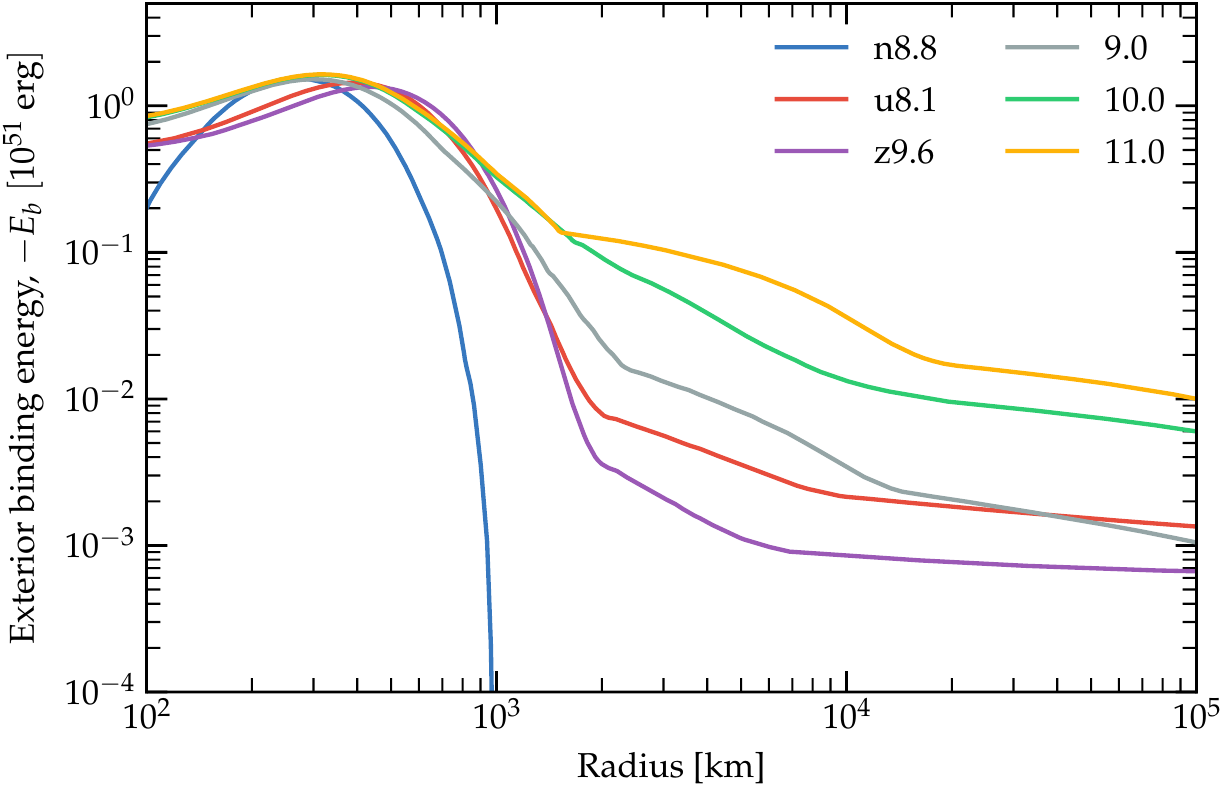}
  \end{minipage}
  \caption{Progenitor models: density profiles in ${\rm g}\, {\rm
  cm}^{-3}$ (\emph{left panel}) and binding energies in Bethes ($1\ {\rm
  B} = 10^{51}\ {\rm erg}$; \emph{right panel}). The envelope binding
  energy is computed as the total energy exterior to a given radius.
  Note that, for numerical reasons, we modified the n8.8 progenitor with
  the addition of a thicker envelope (see the main text for details).
  Progenitors that successfully explode in 1D (n8.8, u8.1, and z9.6)
  have steeper density profiles and smaller binding energies than
  low-mass progenitors that do not explode in 1D.}
  \label{fig:progenitors}
\end{figure*}

The fate of less massive stars in the \ac{ZAMS} range ${\sim} 8\
M_\odot$ to ${\sim} 12\ M_\odot$ is less clear. Depending on the initial
mass, the ultimate fate could be to form massive white dwarfs, to form
iron cores as do regular massive stars, or to explode as \acp{ECSN}
before forming an iron core \citep[e.g.][]{nomoto:1984a, nomoto:1987a,
jones:2013wda, doherty:2015a, woosley:2015a, doherty:2017a}. It has also
been suggested that these stars might undergo violent flashes and power
unusual transients before their deaths \citep{woosley:2015a,
jones:2016asr}.

\acp{ECSN} and low-mass iron-core \acp{CCSN} with similar features are
expected to occur in a relatively narrow range of \ac{ZAMS} masses.
However, they might account for a significant fraction of
gravitational-collapse \acp{SN}, given that the initial mass function of
stars drops rapidly towards high masses. These progenitors have compact
cores with tenuous envelopes, which result in a steep drop of the
accretion rate after core bounce. This, in turn, triggers early
explosions, even under the assumption of spherical symmetry
\citep{kitaura:2005bt, janka:2007di, burrows:2007a, fischer:2009af,
janka:2012sb}. \acp{ECSN} and \ac{ECSN}-like \acp{CCSN} are expected to
be underenergetic and possibly underluminous and to have small
${}^{56}{\rm Ni}$ yields and peculiar nucleosynthetic abundances
\citep[e.g.][]{nomoto:1982a, kitaura:2005bt, janka:2007di,
hoffman:2007du, wanajo:2010ig, melson:2015tia, wanajo:2017cyq}.

\citet{kitaura:2005bt} suggested that \ac{ECSN}-like events might
explain a subclass of Type-II \acp{SN} with unusually low luminosities
\citep{pastorello:2004a, spiro:2014a}. An \ac{ECSN} has also been
invoked to explain SN 1054 and the associated Crab remnant
\citep{nomoto:1982a, takahashi:2013ena, smith:2013gya,
tominaga:2013ala}. According to historical records, SN 1054 was not
underluminous. However, SN-1054 was likely underenergetic, with an
explosion energy around $10^{50}$ erg, as indicated by the low-mass of
the Crab nebula's filaments and their relatively low expansion velocity,
as well as by the small inferred ${}^{56}{\rm Ni}$ yield \citep[and
references therein]{muller:2016izw}. On the basis of measured isotopic
abundance anomalies, it has also been suggested that a low-mass
\ac{CCSN} might have been the trigger that started the formation of our
Solar System \citep{banerjee:2016tzd}.

\acp{ECSN} and \ac{ECSN}-like progenitors have attracted significant
interest in the \ac{CCSN} mechanism community due to their
``explodability'' and the fact that they allow for self-consistent
studies also in 1D (with the assumption of spherical symmetry).
\citet{hillebrandt:1984} performed the first 1D simulations of the
collapse, bounce, and explosion of the original n8.8 \ac{ECSN}
progenitor of Nomoto \citep{nomoto:1984a, nomoto:1987a}, using an
approximate gray neutrino transport scheme. They found an energetic
explosion by the prompt shock mechanism with an energy of ${\sim}2\cdot
10^{51}$ erg. Subsequent studies, with modern neutrino interactions and
multi-group transport in 1D and 2D, performed by \citet{kitaura:2005bt,
janka:2007di, burrows:2007a} and \citet{fischer:2009af} found much
weaker explosions (${\sim} 10^{50}\ {\rm erg}$) powered by the delayed
neutrino mechanism. \citet{mueller:2012ak} considered an $8.1\ M_\odot$
(\ac{ZAMS}) progenitor with metallicity $Z = 10^{-4}$, u8.1, which
formed an iron core, but had a stellar structure very similar to the
n8.8 progenitor, and found a similarly early explosion. Another
iron-core progenitor, the zero-metallicity $9.6$-$M_\odot$ model, z9.6,
was considered and determined to have a qualitatively similar outcome by
\citet{janka:2012sb, mueller:2012sv, mueller:2014rna} in 2D, and by
\citet{melson:2015tia} in 3D. Very recently, \citet{wanajo:2017cyq}
presented new nucleosynthetic calculations for the n8.8, u8.1, and z9.6
models, as well as a summary of their associated explosion
characteristics with the \textsc{Coconut}-\textsc{Vertex} code.

The low-mass, but otherwise ``canonical,'' $11.2$-$M_\odot$ progenitor
from \citet{woosley:2002zz} and the $12.0$-$M_\odot$ progenitor from
\cite{woosley:2007as} have been considered by several groups,
\citep[e.g.][]{buras:2005tb, takiwaki:2011db, mueller:2012is,
bruenn:2013a, bruenn:2014qea, dolence:2014rwa, muller:2015dia,
summa:2015nyk, oconnor:2015rwy, burrows:2016ohd, nagakura:2017mnp}.
While the $11.2$-$M_\odot$ progenitor has a post-bounce evolution that
is qualitatively similar to the u8.1 progenitor \citep{mueller:2012ak},
the $12.0$-$M_\odot$ progenitor either explodes very late
\citep{summa:2015nyk} or not at all \citep{oconnor:2015rwy,
dolence:2014rwa, skinner:2015uhw}, with the exception of the simulations
by \citet{bruenn:2013a, bruenn:2014qea}. This challenges the commonly
held idea that low-mass iron-core progenitors should explode easily and
similarly to an \ac{ECSN}.

\begin{table*}
\caption{Details of the setup for the perturbed models.}
\label{tab:perturb}
\vspace{-1em}
\begin{center}
\begin{tabular}{lccccccccccccccc}
\toprule
Prog. &
$r_1^-$ & $r_1^+$ & $\ell_1$ & $n_1$ & $\delta v_1$ &
$r_2^-$ & $r_2^+$ & $\ell_2$ & $n_2$ & $\delta v_2$ &
$r_3^-$ & $r_3^+$ & $\ell_3$ & $n_3$ & $\delta v_3$ \\
&
[km] & [km] & & & $[10^8\ {\rm cm}\ {\rm s}^{-1}]$ &
[km] & [km] & & & $[10^8\ {\rm cm}\ {\rm s}^{-1}]$ &
[km] & [km] & & & $[10^8\ {\rm cm}\ {\rm s}^{-1}]$ \\
\midrule
\phantom{0}9.0 &
850   & 1,350    & 12 & 1 & 1.0 &
2,500  & 9,750   & 10 & 1 & 0.3 &
15,000 & 450,000 & 4  & 1 & 0.3 \\
10.0 &
480  & \phantom{0,}560  & 12 & 1 & 1.0 &
1,150 & 1,800 & 10  & 3 & 1.0 &
\phantom{0}2,500 & \phantom{00}3,400 & 4  & 1 & 0.5 \\
11.0 &
450  & \phantom{0,}520   & 12 & 1 & 0.5 &
1,100 & 1,400  & 10  & 1 & 1.2 &
\phantom{0}1,550 & \phantom{0}11,000 & 4  & 1 & 0.5 \\
\bottomrule
\end{tabular}
\end{center}
\end{table*}

With the goal of characterizing the different explosions of \acp{ECSN},
\ac{ECSN}-like \ac{CCSN}, and canonical, but low-mass, \acp{CCSN}, we
present 1D and 2D \fornax simulations (Dolence et al.  2017, in prep.)
of the collapse, bounce, and subsequent evolution of six progenitor
models. We consider the \ac{ECSN} n8.8 progenitor from
\citet{nomoto:1984a, nomoto:1987a}, the \ac{ECSN}-like u8.1 and z9.6
progenitors from Heger (private communication), and the 9-, 10-, and
$11$-$M_\odot$ solar-metallicity iron-core collapse progenitors from
\citet{sukhbold:2016a}. We show that low-mass \acp{CCSN} always fail to
explode in 1D and evolve in a qualitatively different way compared to
\acp{ECSN} and \ac{ECSN}-like \acp{CCSN}. We also discuss the impact of
the dimensionality, of pre-supernova perturbations, and of changes in
the microphysics. In particular, for the latter, we focus on the effects
of many-body corrections to the axial-vector term in the
neutrino-nucleon scattering rate recently explored by
\citet{horowitz:2016gul}.

The rest of this paper is organized as follows. First, in Section
\ref{sec:methods}, we give an overview of the simulation setup and of
the properties of the progenitor models. We discuss the qualitative
outcome of our simulations in Section \ref{sec:overall}, while a more
quantitative account of the energetics of the explosions is given in
Section \ref{sec:explene}. We discuss the properties of the neutrino
radiation in Section \ref{sec:nurad}. Section \ref{sec:pns} is dedicated
to the properties and evolution of the remnant \acp{PNS}. Finally, we
summarize and discuss our results in Section \ref{sec:conclusions}.

\section{Progenitors and Setup}
\label{sec:methods}

As previously discussed, we consider six progenitor models, which we
label as n8.8 \citep{nomoto:1984a, nomoto:1987a}, u8.1 and z9.6 (Heger
private communication), and 9.0-, \mbox{10.0-,} $11.0$-$M_\odot$
\citep{sukhbold:2016a}. The u8.1 progenitor has metallicity $10^{-4}$ of
Solar, the z9.6 has zero metallicity, and all other progenitors have
Solar metallicity. All of the progenitors have been evolved up to
the point of core-collapse, defined as the time their radial infall
velocity has reached ${\sim}1000$ ${\rm km}\ {\rm s}^{-1}$. Their
structure (density and exterior binding energies) are shown in
Fig.~\ref{fig:progenitors}.  All of these progenitors have relatively
compact cores and loosely bound envelopes. The values of the compactness
parameter $\xi_{2.5}$ \citep{oconnor:2010moj, oconnor:2012bsj} computed
from the progenitor models are given in Tab.~\ref{tab:main}. They range
from $\simeq 7.6\cdot 10^{-5}$, for the z9.6 progenitor, to $\simeq
7.7\cdot 10^{-3}$, for the 11.0-$M_\odot$ progenitor.  $\xi_{2.5}$
cannot be computed for the n8.8 progenitor since it only extends to
$\simeq 1.32\ M_\odot$. Note that, for numerical reasons, we modify the
n8.8 progenitor for $\rho \leq 10^{4}\ {\rm g}\cdot {\rm cm}^{-3}$ with
the addition of a constant temperature envelope with $\rho \propto
r^{-2}$.  We verified, by changing the density of the envelope over 3
orders of magnitude, that the explosion energy of the n8.8 progenitor is
not very sensitive to this envelope, although the shock propagation speed is
obviously affected.

All of the progenitors are non-rotating and have been evolved in 1D.
There is currently significant interest in the possible impact of
pre-supernova turbulence on the development of explosions, which several
authors have determined to be beneficial and, in some cases, crucial to
the outcome \citep{couch:2013coa, mueller:2014oia, abdikamalov:2016bzh,
takahashi:2016hdc, muller:2017hht}. Of relevance, the first progenitor
models evolved in 3D shortly before core collapse have also recently
become available \citep{couch:2015gua, mueller:2016a}. In this spirit,
we also consider the impact of perturbations on the 9.0-, 10.0-, and
11.0-$M_\odot$ progenitors using the approach introduced by
\citet{mueller:2014oia}, which we briefly describe. We introduce
velocity perturbations in three regions $r_i^- \leq r \leq r_i^+$,
$i=1,2,3$, obeying the divergence-free condition
$\nabla\cdot(\rho\,\delta\mathbf{v}_i)=0$. These are generated as
\begin{equation}
  \delta\mathbf{v}_i = \left\{\begin{array}{ll}
    \frac{C_i}{\rho}
      \nabla\times\bm{\Psi}_i, & r_i^- \leq r \leq r_i^+, \\
    0, & \textrm{otherwise};
  \end{array}\right.
\end{equation}
where
\begin{equation}
  \bm{\Psi}_i = \mathbf{e}_\phi \frac{\sqrt{\sin\theta}}{r}
  \sin\left( n_i \pi \frac{r - r_i^-}{r_i^+ - r_i^-}\right ) Y_{\ell_{i},
  1}(\theta, 0)
\end{equation}
and $n_i$, $\ell_i$ are the number of convective cells in the radial
and angular directions, respectively. Finally, $C_i$ is tuned to achieve
a given maximum perturbation amplitude. The parameters we use are given
in Tab.~\ref{tab:perturb}.

We evolve these progenitors from the onset of collapse with the
neutrino-radiation-hydrodynamics code \mbox{\fornax} (\citealt{skinner:2015uhw,
burrows:2016ohd}; Dolence et al.~2017, in prep.).  \fornax solves for
the transport of neutrinos using a multi-dimensional moment scheme with
an analytic closure for the \nth{2} and \nth{3} moments
\citep{shibata:2011kx, murchikova:2017zsy}. Similar moment methods have
also been recently adopted for the \ac{CCSN} problem by other groups
\citep[\eg][]{oconnor:2014sgn, just:2015fda, oconnor:2015rwy,
roberts:2016lzn}. The moment equations are solved using a \nth{2}-order
finite-volume scheme with the HLLE approximate Riemann solver
\citep{einfeldt:1988a}, modified as in \citet{audit:2002hr} and
\citet{oconnor:2014sgn} to reduce the numerical dissipation in the
diffusive limit. \fornax separately evolves electron neutrinos $\nu_e$
and anti-electron neutrinos $\bar{\nu}_e$, while heavy-lepton neutrinos
$\nu_{\mu}$, $\nu_{\tau}$, and the respective anti-particles are lumped
together as a single species, which we denote as ``$\nu_\mu$'' (see
\citealt{bollig:2017lki} for a discussion of the possible limitation of
this aporach). The energy spectra of neutrinos are resolved using 20
logarithmically-spaced energy groups extending to $300\ {\rm MeV}$ for
electron neutrinos and to $100\ {\rm MeV}$ for anti-electron and
heavy-lepton neutrinos.

\begin{figure*}
  \begin{minipage}{\columnwidth}
    \includegraphics[width=\columnwidth]{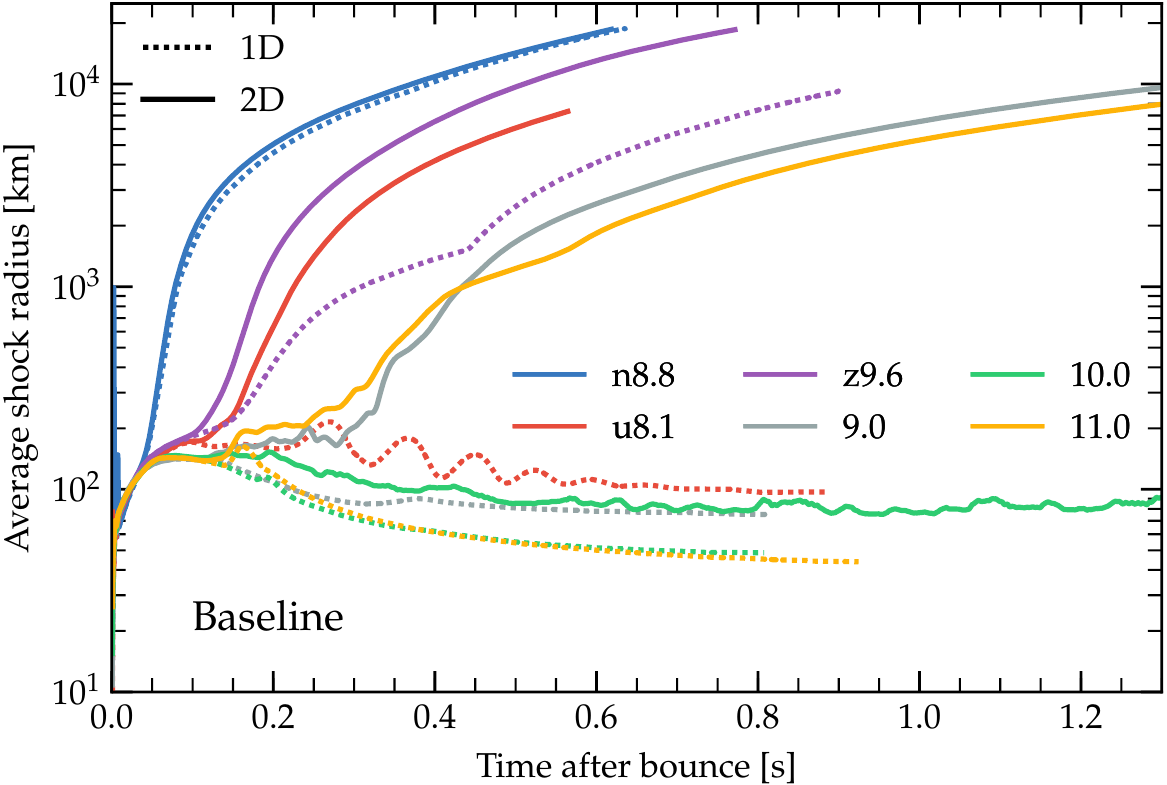}
  \end{minipage}
  \hfill
  \begin{minipage}{\columnwidth}
    \includegraphics[width=\columnwidth]{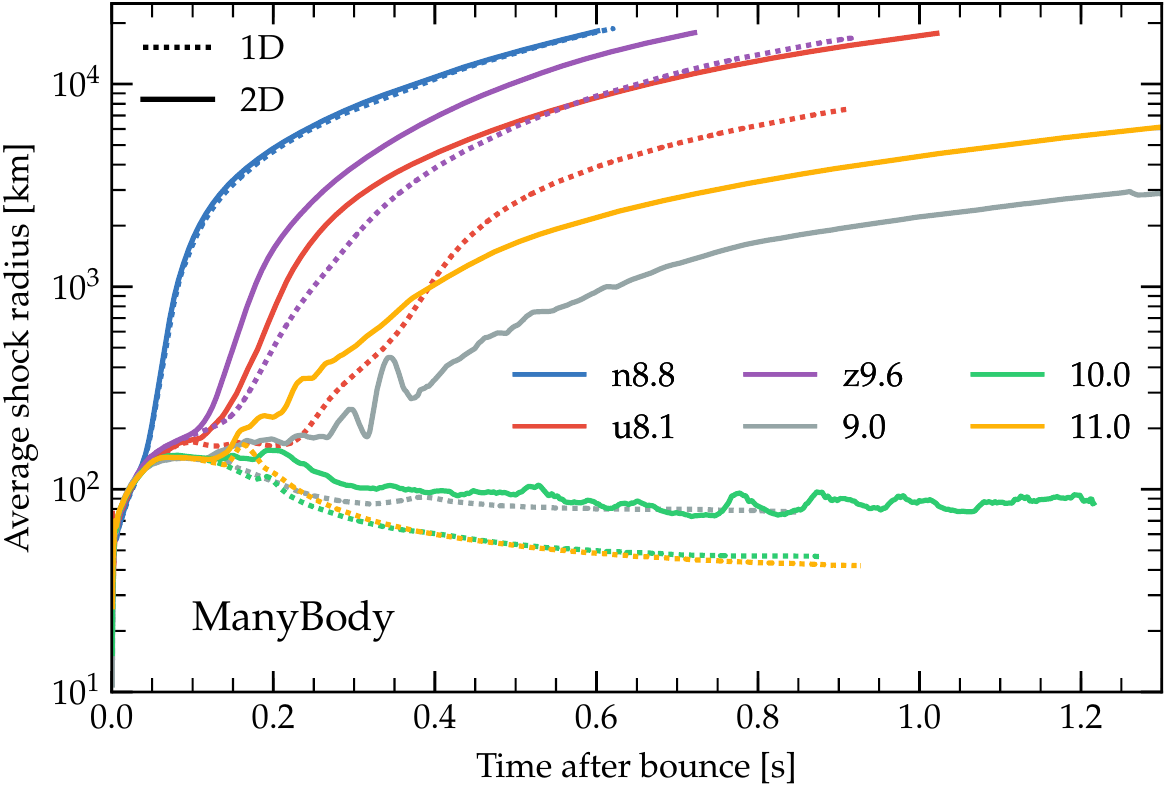}
  \end{minipage}
  \caption{Average shock radius (km) tracks for all progenitors in 1D
  and 2D with our Baseline setup (\emph{left panel}) and with the
  inclusion of many-body corrections (\emph{right panel}). The curves
  are smoothed using a running average with a 5-ms window. With
  many-body corrections the u8.1 progenitor explodes also in 1D, and the
  9.0- and 11.0-$M_\odot$ 2D explosions become more robust. None of the
  progenitors from Sukhbold et al.  (2016) explode in 1D, even with
  many-body corrections. The 10.0-$M_\odot$ only explodes when both
  many-body corrections and perturbations are included.}
  \label{fig:rshock}
\end{figure*}

\begin{figure}
  \includegraphics[width=\columnwidth]{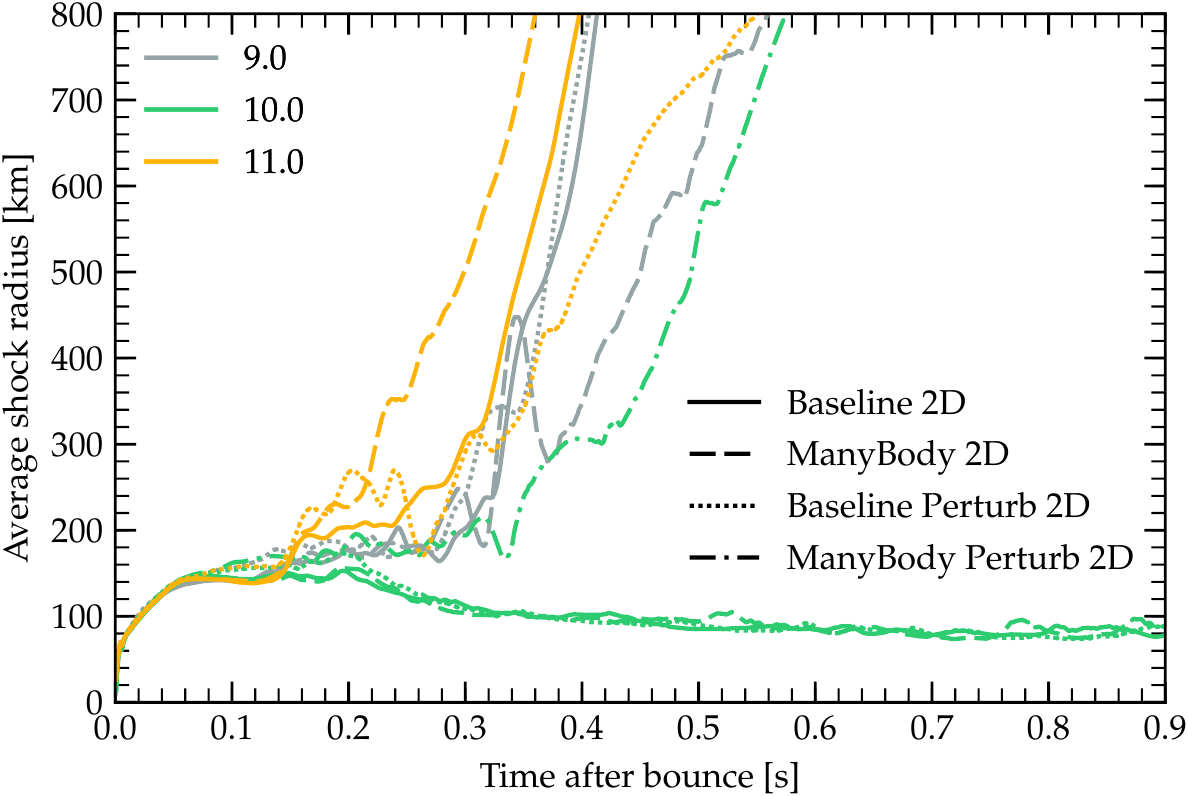}
  \caption{Impact of perturbations and changes to the microphysics on
  the average shock radius (km) of the 9.0-, 10.0-, and 11.0-$M_\odot$
  progenitors. The curves are smoothed using a running average with a
  5-ms window. Both the inclusion of perturbations and of many-body
  effects can have a qualitative and quantitative impact.}
  \label{fig:rshock.sukhbold}
\end{figure}

The set of neutrino-matter interactions included in our simulations are
described \citet{burrows:2004vq}. We include weak magnetism and recoil
correction to neutrino-nucleon scattering and absorption
\citep{horowitz:2001xf}. We treat inelastic neutrino-electron scattering
with the scheme of \citet{thompson:2002mw} and the relativistic
formalism summarized in \citet{reddy:1998hb}. Inelastic neutrino-nucleon
scattering is included using the formalism of \citet{thompson:2002mw},
for $\nu_e$ and $\bar{\nu}_e$, while we use the approach of
\citet{mueller:2014oia} for inelastic scattering of heavy-lepton
neutrinos on nucleons. For the latter, we use $6 k_{\rm B} T$, instead
of $3 k_{\rm B} T$ for the crossover energy between upscattering and
downscattering \citep{thompson:2000gv, tubbs:1979a}, where $k_{\rm B}$
is Boltzmann's constant and $T$ is the temperature.  That is, we
approximate the redistribution rate of the heavy-lepton neutrinos to be
proportional to $( \epsilon_{\nu} - 6  k_{\rm B} T)/m_nc^2$, where $m_n$
is the neutron mass and $c$ is the speed of light, and $\epsilon_\nu$ is
the incoming neutrino energy. Electron capture on heavy nuclei during
the infall is treated following \citet{bruenn:1985en}. However, we
disable electron capture on heavy nuclei for the n8.8 progenitor to
prevent an unphysical neutronization burst during the infall, which is
due to the failure of our nuclear statistical equilibrium approximation
in the outer core of this progenitor.

We perform two variants of each simulation. ``Baseline'' includes all
the neutrino-matter interaction discussed above. ``ManyBody'' also
include many-body corrections to the neutrino-nucleon scattering
cross-section as estimated by \citet{horowitz:2016gul}. These are
implemented using the fit to the axial response factor $S_A$ they
provided. The runs with perturbations are performed using the Baseline
physics setup, with the exception the 10.0-$M_\odot$ which is evolved
with both the Baseline and the ManyBody setup.

The hydrodynamic equations are solved using a high-resolution
shock-capturing scheme with \nth{3}-order reconstruction and the HLLC
approximate Riemann solver \citep{toro:1994a}. The details of the
numerical schemes are discussed in (\citealt{skinner:2015uhw,
burrows:2016ohd}; Dolence et al.~2017, in prep.).  For the simulations
presented here, we use a spherical grid with 678 points extending up to
20,000 km.  The grid has a constant spacing $\Delta r$ of 0.5 km for $r
\lesssim 10\ {\rm km}$ and then smoothly transitions to a
logarithmically spaced grid with $\Delta r/r \simeq 0.01$ for $r \gtrsim
100\ {\rm km}$. For the 2D simulations, we use 256 angular zones with
angular resolution smoothly varying between $\simeq 0.95^\circ$ at the
poles and $\simeq 0.64^\circ$ at the equator. The angular grid is also
progressively derefined towards the center, \ie we use a dendritic grid
(Dolence et al.~2017, in prep.), to avoid an excessively restrictive CFL
condition in the angular direction.

We adopt the Lattimer-Swesty equation of state with nuclear
compressibility parameter $220\ {\rm MeV}$ \citep{lattimer:1991nc} and
treat gravity in the monopole approximation using a general-relativistic
potential, following \citet{marek:2005if}.

Finally, we carry out all 2D simulations until the maximum shock radius
exceeds 19,000 km, or until the explosion is deemed unsuccessful.

\section{Overall Dynamics}
\label{sec:overall}

A first glance of our results can be gained from Fig.~\ref{fig:rshock},
which shows the average shock radii for all progenitors in 1D and 2D
with both the Baseline and ManyBody setups. As in previous works by
others \citep{kitaura:2005bt, janka:2007di, burrows:2007a,
fischer:2009af, mueller:2012ak, janka:2012sb, mueller:2012sv,
melson:2015tia}, we find early explosions for the n8.8, z9.6, and u8.1
progenitors. The same progenitors also explode in 1D spherical symmetry,
although the u8.1 only with the ManyBody setup.

In the n8.8 models, the shock suddenly accelerates outwards starting
from ${\sim} 50\ {\rm ms}$ after bounce, a few tens of milliseconds
earlier than in \citet{kitaura:2005bt} and \citet{janka:2007di}, but
similarly to \citet{fischer:2009af}. We note that these works used
different equations of state compared to us.  \citet{kitaura:2005bt} and
\citet{janka:2007di} used the Lattimer-Swesty equation of state with
nuclear compressibility parameter $180\ {\rm MeV}$
\citep{lattimer:1991nc}, while \citet{fischer:2009af} used the equation
of state of \citet{shen:1998by}. These studies also differ in their
treatment of nuclear burning and electron capture during infall. These
differences result in variations of the positions at which the shock
originally forms, which in turn might explain the different explosion
times. In the simulations of the Garching group the homologous core at
bounce has a mass of $0.425\ M_\odot$ \citep{kitaura:2005bt}, while
\citet{fischer:2009af} report a larger core mass of $0.625\ M_\odot$.
In our simulations, the homologous core encloses a mass of $0.611\
M_\odot$ at bounce.

\begin{figure}
  \includegraphics[width=\columnwidth]{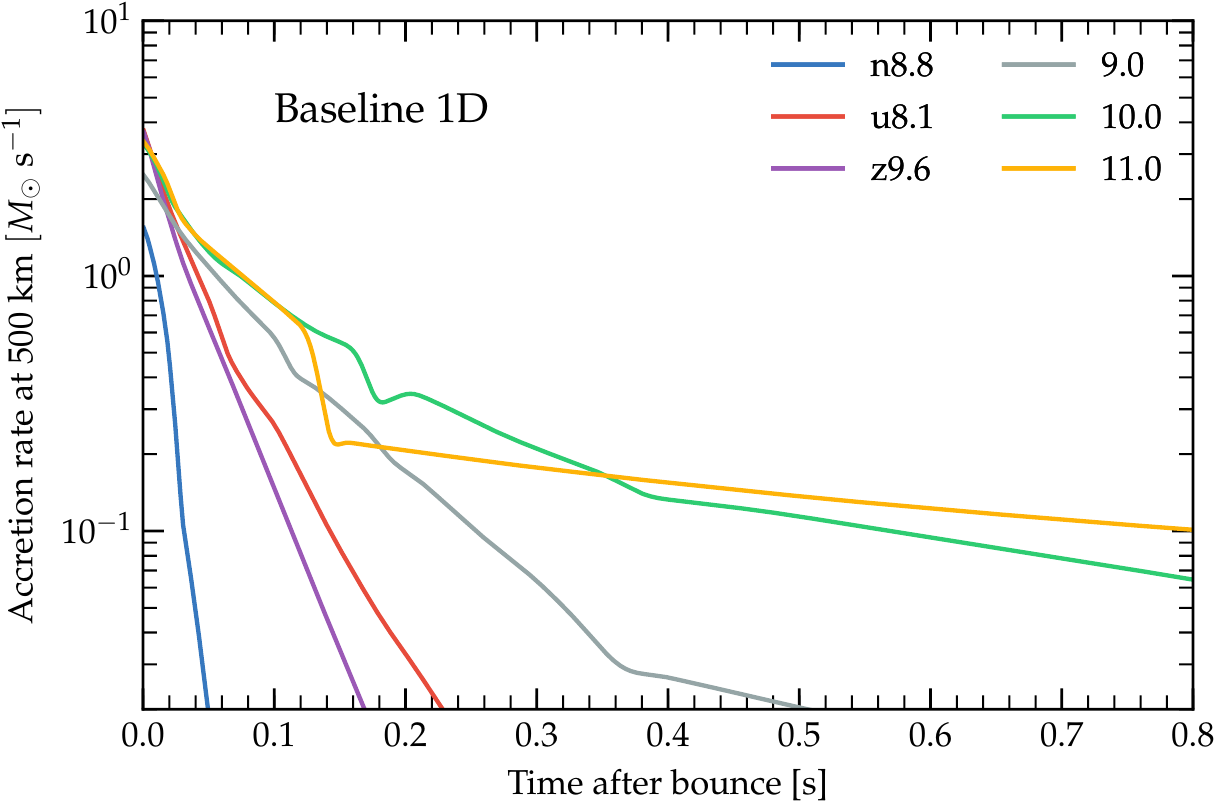}
  \caption{Accretion rates in $M_\odot\ {\rm s}^{-1}$ at 500 km for our
  Baseline 1D setup. The curves are smoothed using a running average
  with a 5-ms window. Successful 1D explosions require steep drops in
  the accretion rate at early times, when the neutrino luminosities are
  still large.}
  \label{fig:mdot}
\end{figure}

None of the progenitors from \citet{sukhbold:2016a} explode in
self-consistent 1D simulations. Somewhat surprisingly, the
10.0-$M_\odot$ progenitor fails to explode, within the simulation time,
also in 2D. However, both the 9.0- and 11.0-$M_\odot$ progenitor explode
successfully in 2D, with either the Baseline or the ManyBody setups.

\begin{figure*}
  \includegraphics[width=\textwidth]{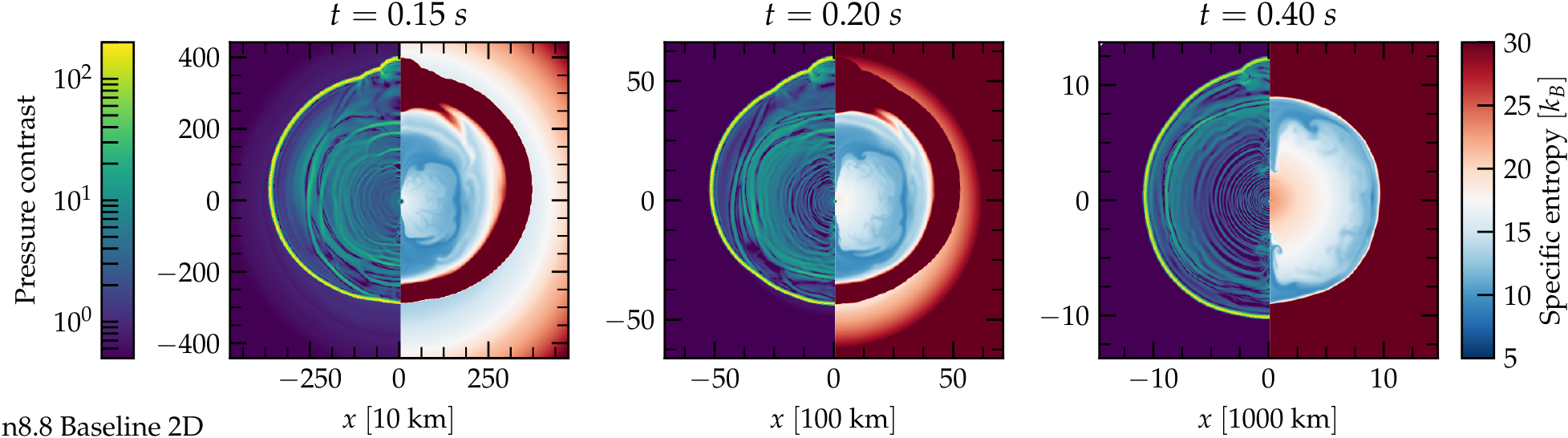}
  \\[1em]
  \includegraphics[width=\textwidth]{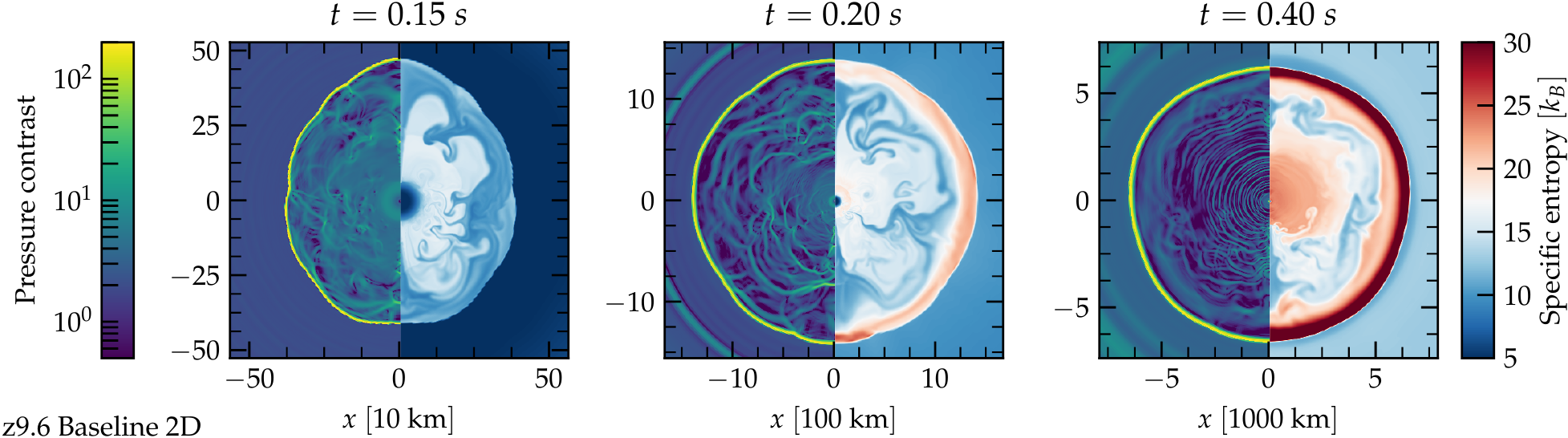}
  \\[1em]
  \includegraphics[width=\textwidth]{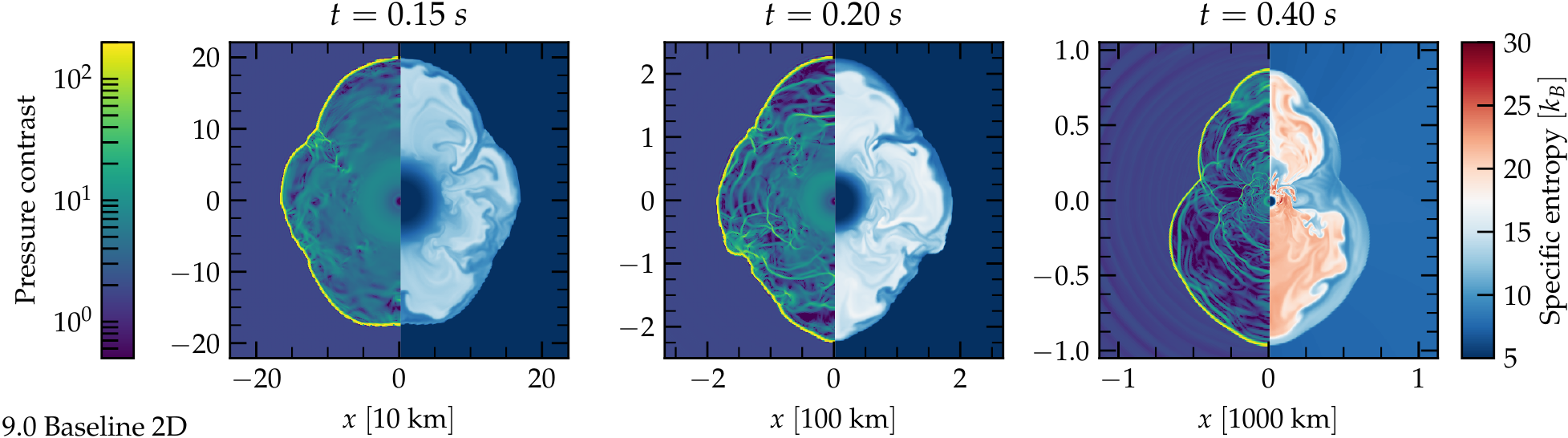}
  \\[1em]
  \includegraphics[width=\textwidth]{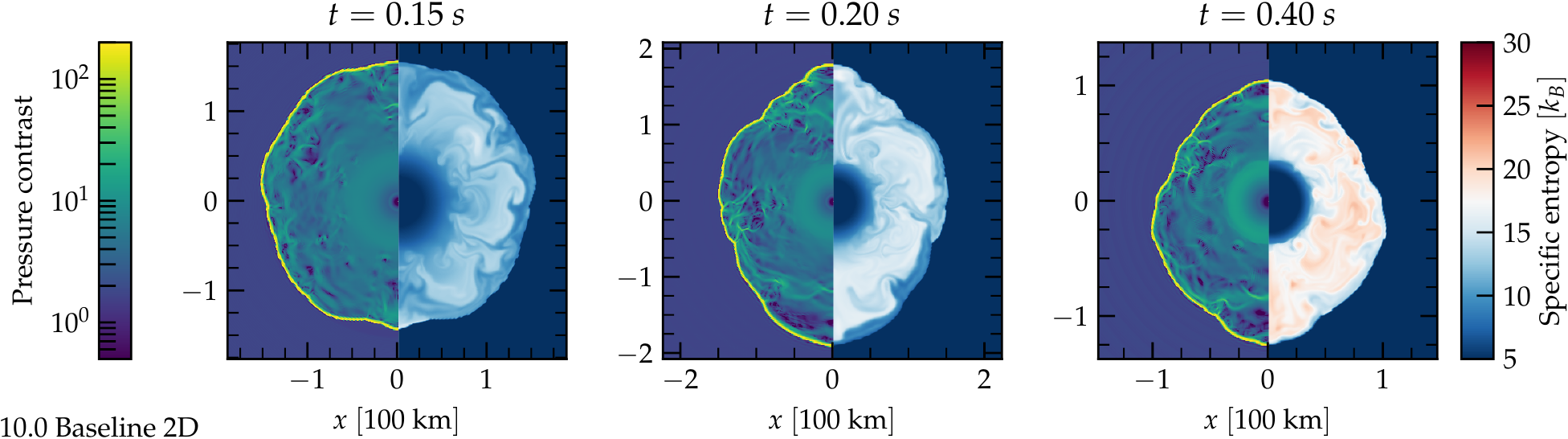}
  \caption{Entropy per baryon in $k_B$ and pressure contrast
  ($r |\nabla p|/p$) profiles for the n8.8, z9.6, 9.0, and 10.0
  progenitors evolved with the Baseline setup at three representative
  times. Note the different spatial scales. The ring-like structure
  visible in the pressure contrast in some panels are compositional
  shells. All models develop convection around ${\sim} 0.15\ {\rm s}$
  after bounce. The z9.6 model shows a nearly-symmetrical explosion,
  while the 9.0-$M_\odot$ progenitor develops asymmetric explosions. The
  10.0-$M_\odot$ progenitor does not explode with the Baseline setup.}
  \label{fig:entropy2d.baseline}
\end{figure*}

\begin{figure*}
  \includegraphics[width=\textwidth]{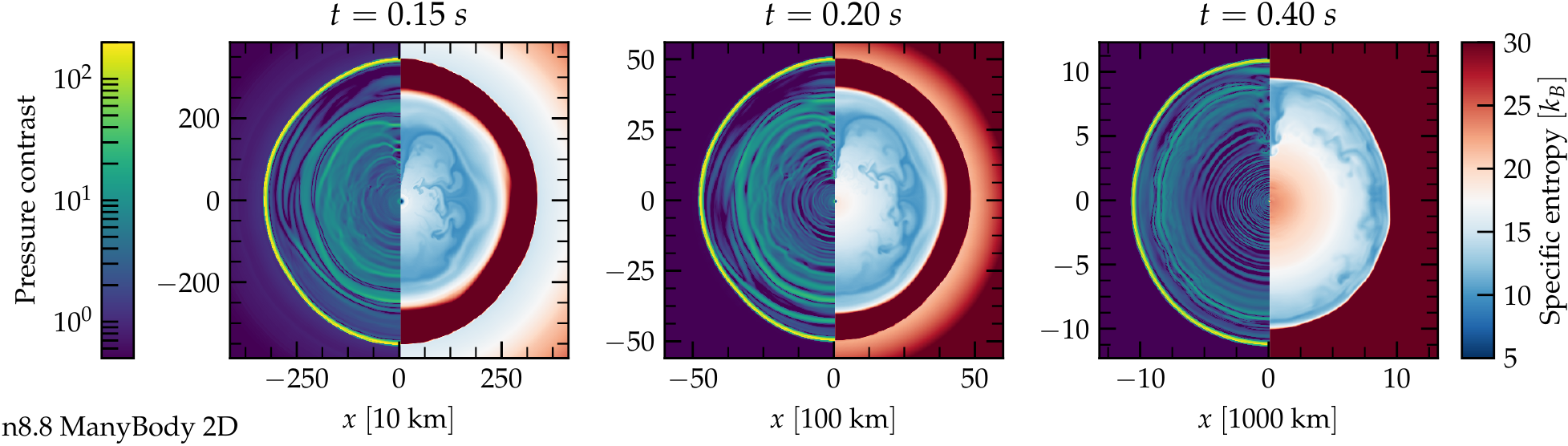}
  \\[1em]
  \includegraphics[width=\textwidth]{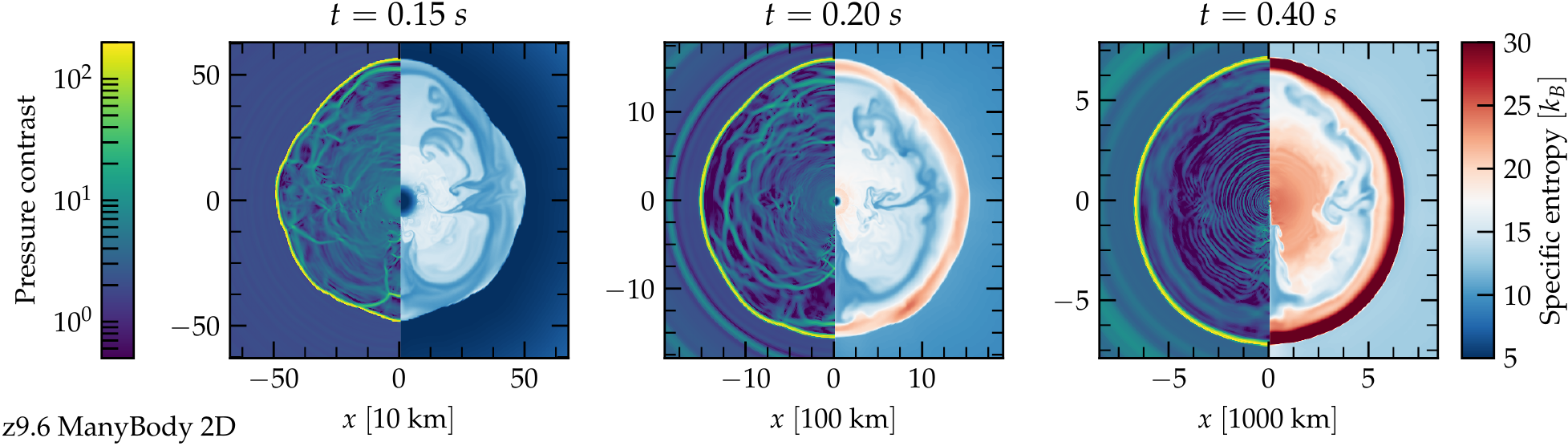}
  \\[1em]
  \includegraphics[width=\textwidth]{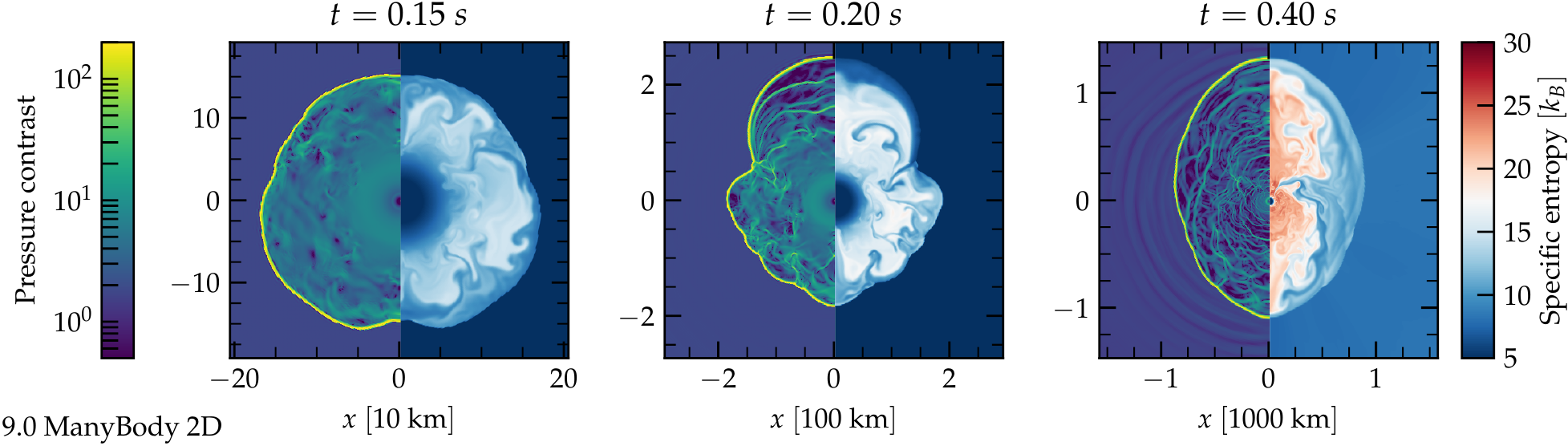}
  \\[1em]
  \includegraphics[width=\textwidth]{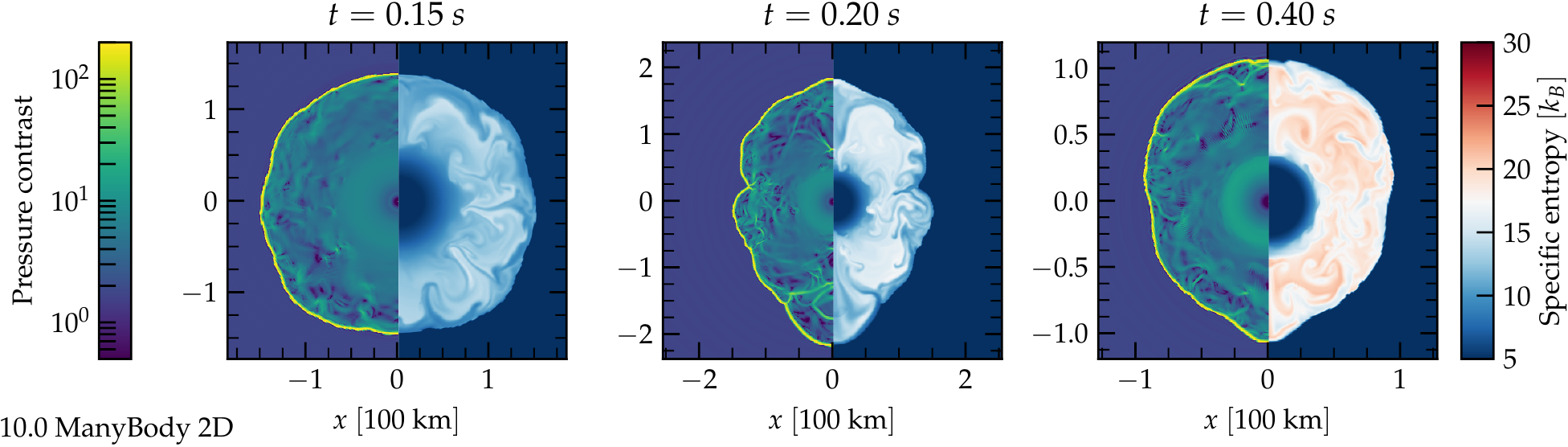}
  \caption{Entropy per baryon in $k_B$ and pressure contrast
  ($r |\nabla p|/p$) profiles for the n8.8, z9.6, 9.0, and 10.0
  progenitors evolved with many-body corrections at three representative
  times. This figure should be contrasted with
  Fig.~\ref{fig:entropy2d.baseline}. Compared to the Baseline setup, the
  inclusion of many-body corrections results in larger entropies and
  more violent convective overturn at early times. The many-body
  corrections included in our ManyBody setup are not sufficient to turn
  the 10.0-$M_\odot$ from a dud to a successful explosion. However, they
  yield more symmetric explosions for the 9.0-$M_\odot$ progenitor.}
  \label{fig:entropy2d.horowitz}
\end{figure*}

\begin{figure*}
  \begin{minipage}{\columnwidth}
    \includegraphics[width=\columnwidth]{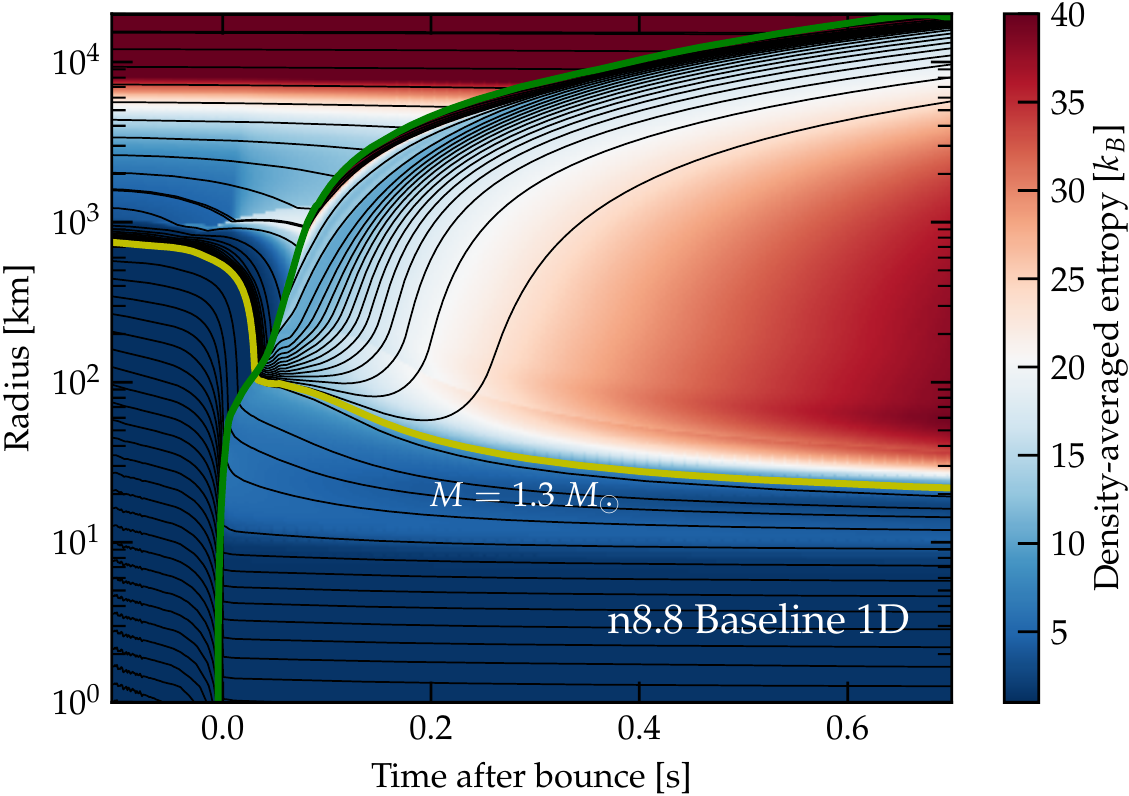}
  \end{minipage}
  \hfill
  \begin{minipage}{\columnwidth}
    \includegraphics[width=\columnwidth]{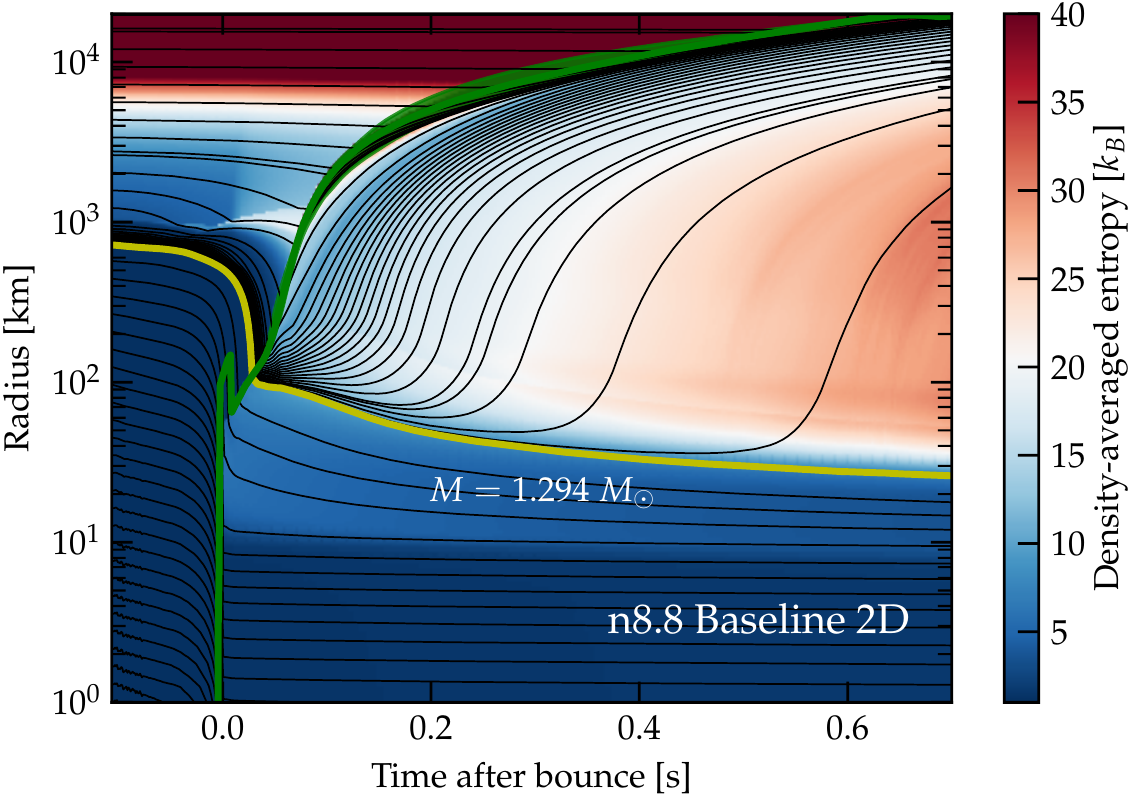}
  \end{minipage}
  \caption{Evolution of the n8.8 progenitor in 1D (\emph{left panel})
  and 2D (\emph{right panel}) with the Baseline setup. The green line
  denotes the average shock radius. The black lines are curves of
  constant enclosed baryonic mass (Lagrangian fluid elements in 1D). The
  yellow thick line denotes the final PNS mass cut. The curves are smoothed
  using a running average with a 5-ms window. The background color is
  the density-averaged entropy per baryon in $k_B$. 1D explosions
  generically result in the creation of low-density, high-entropy
  bubbles, which are smeared out by convection in 2D.}
  \label{fig:summary.n8p8}
\end{figure*}

\begin{figure*}
  \begin{minipage}{\columnwidth}
    \includegraphics[width=\columnwidth]{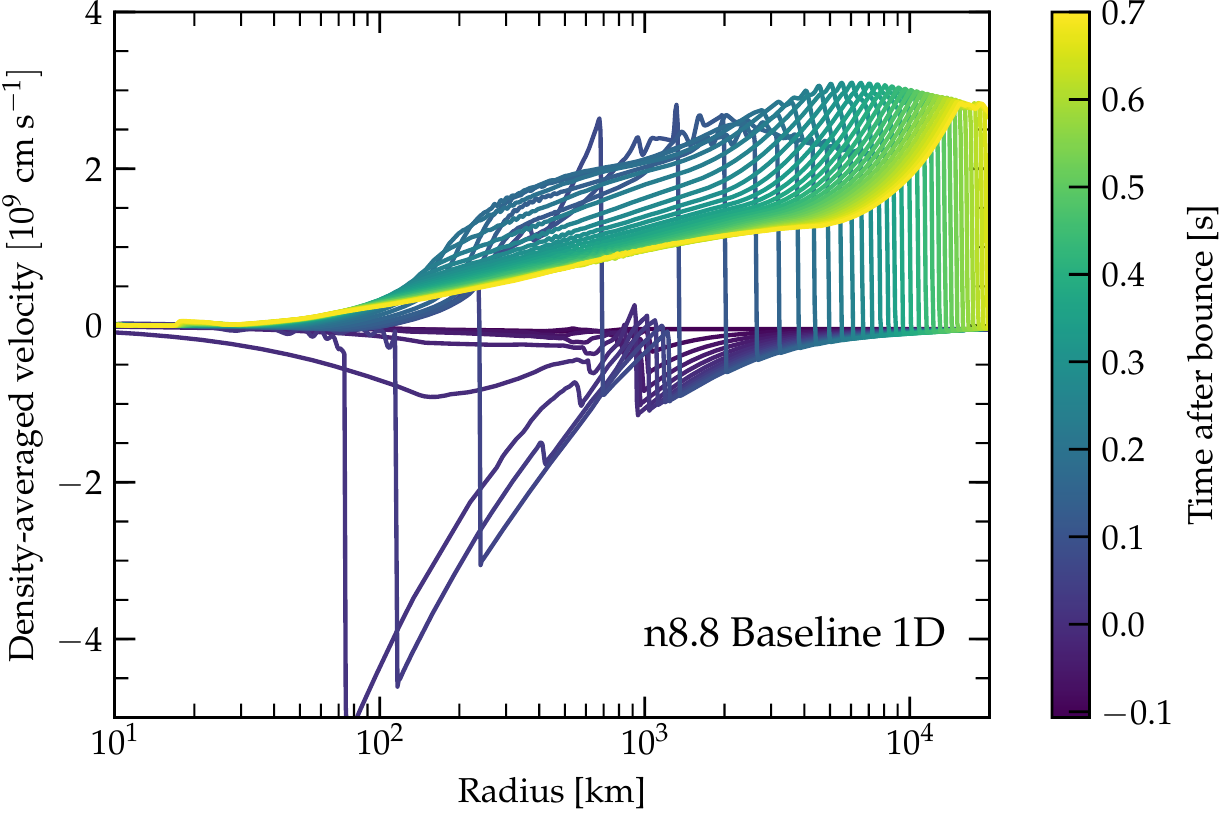}
  \end{minipage}
  \hfill
  \begin{minipage}{\columnwidth}
    \includegraphics[width=\columnwidth]{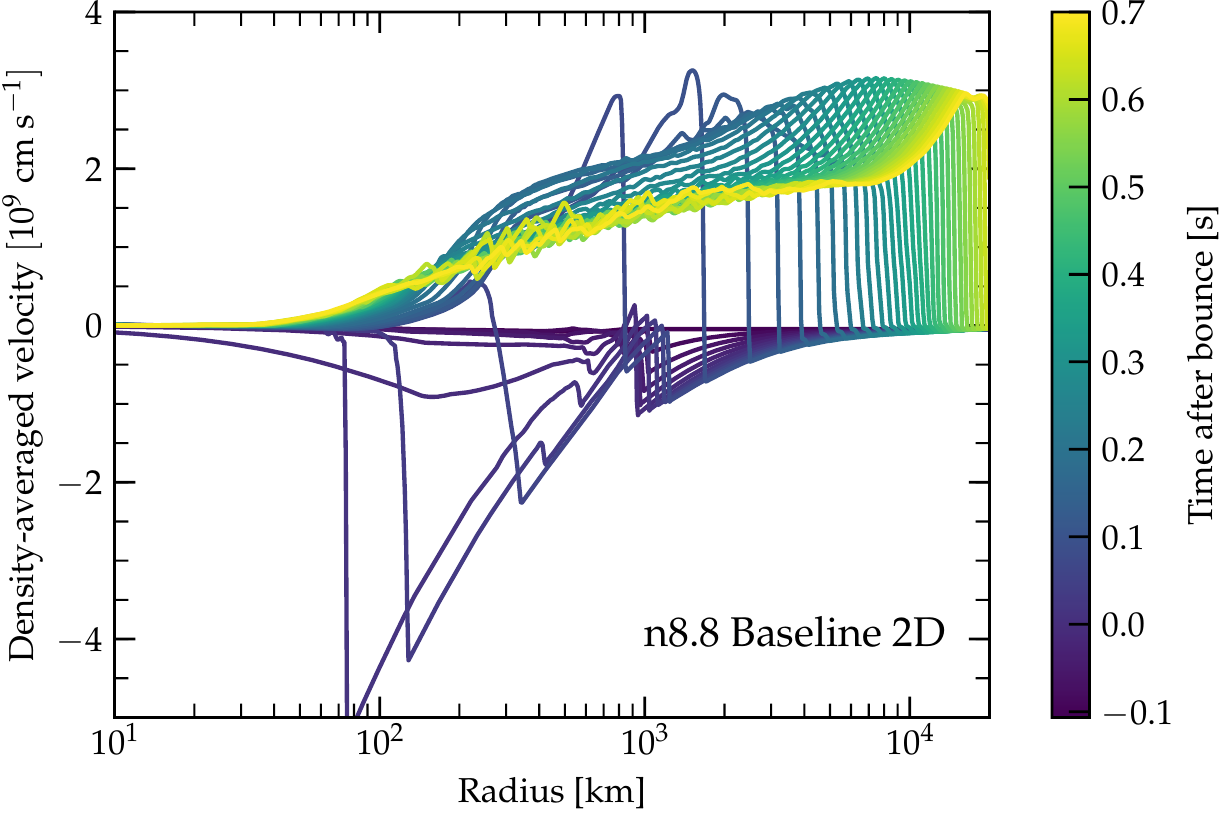}
  \end{minipage}
  \caption{Density-averaged radial velocity in units of $10^9\ {\rm cm}\
  {\rm s}^{-1}$ for the n8.8 progenitor evolved with Baseline physics in
  1D (\emph{left panel}) and 2D (\emph{right panel}). Multi-dimensional
  explosions result in larger velocities (and kinetic energies) in the
  neutrino driven wind.}
  \label{fig:velr.n8p8}
\end{figure*}

To visualize the different outcomes of the progenitors from
\citet{sukhbold:2016a} in 2D, we highlight their early post-bounce
average shock radii in Fig.~\ref{fig:rshock.sukhbold}. The 9.0-$M_\odot$
and 11.0-$M_\odot$ models have delayed explosions at ${\sim}0.3\ {\rm
s}$ after bounce (Baseline setup) or ${\sim}0.2\ {\rm s}$ after bounce
(11.0-$M_\odot$, with ManyBody). The inclusion of perturbations does not
affect the outcome of the 9.0-$M_\odot$ progenitor. Surprisingly,
perturbations result in a somewhat weaker explosion for the
11.0-$M_\odot$ progenitor, as can be inferred from the smaller shock
expansion velocity and as quantified in Sec.~\ref{sec:explene}. In the
case of the 10.0-$M_\odot$ progenitor, perturbations are able to trigger
a weak explosion ${\sim}0.4\ {\rm s}$ after bounce, but only in
combination with the many-body effects included in the ManyBody setup.
This is the only model for which we find perturbations to yield a
qualitative change to the evolution, despite the fact that the amplitude
of the initial perturbations is at the upper end of what could be
considered as realistic, with turbulent velocities reaching ${\sim}1000\
{\rm km}\ {\rm s}^{-1}$. 

\begin{figure*}
  \begin{minipage}{\columnwidth}
    \includegraphics[width=\columnwidth]{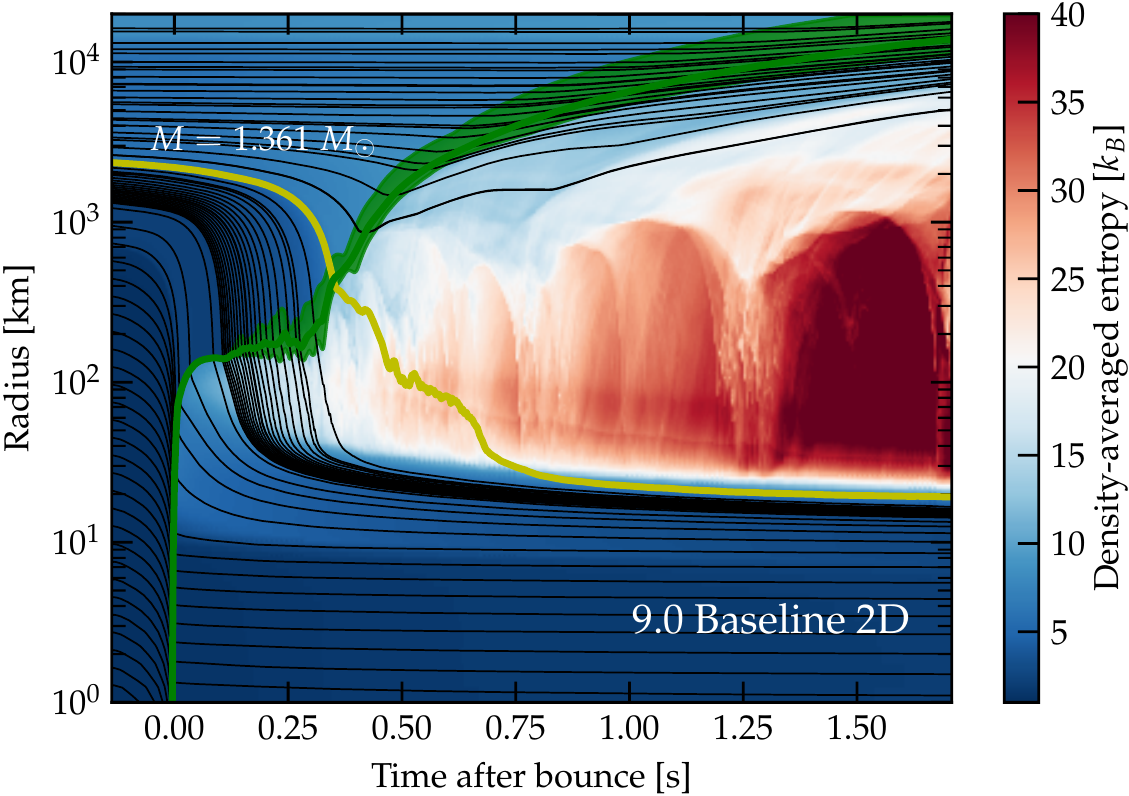}
  \end{minipage}
  \hfill
  \begin{minipage}{\columnwidth}
    \includegraphics[width=\columnwidth]{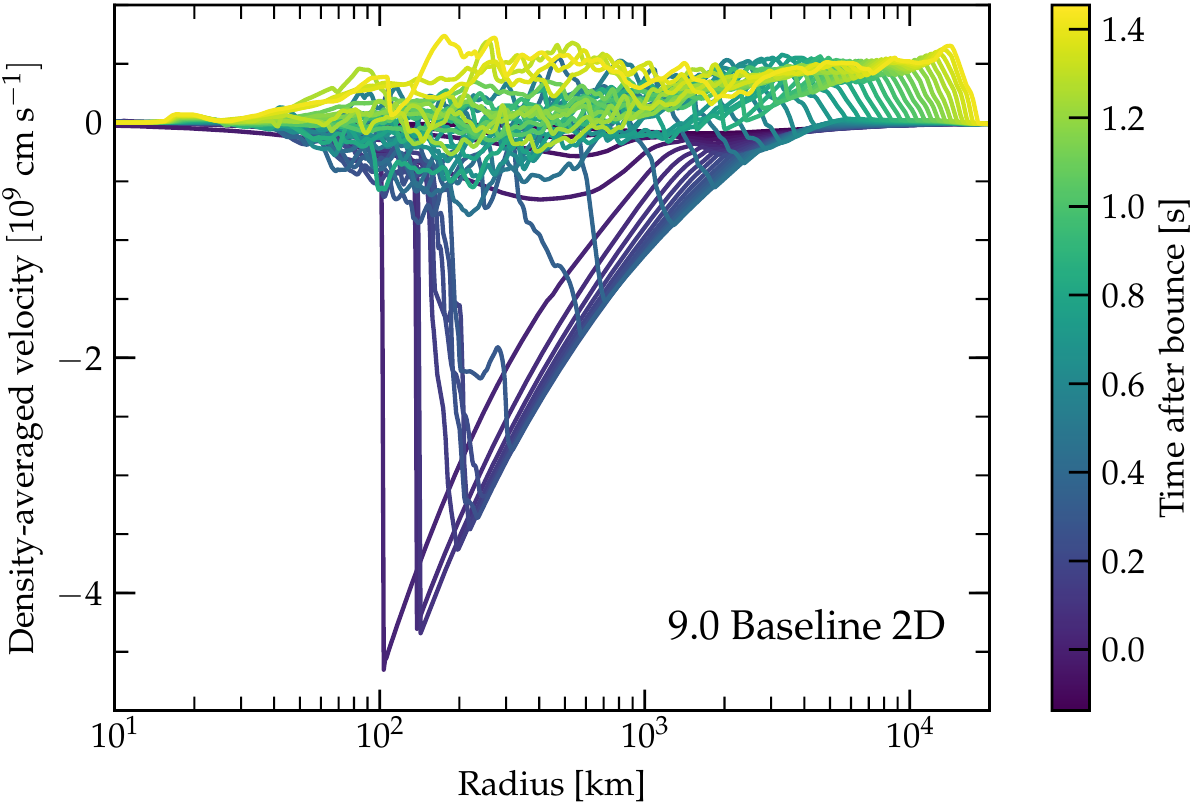}
  \end{minipage}
  \caption{Evolution summary (\emph{left panel}) and density-averaged
  velocity (\emph{right panel}) for the 9.0-$M_\odot$ progenitor with
  the Baseline setup in 2D. The green shaded region in the left panel
  denotes the minimum and maximum shock radius. Curves in the left panel
  are smoothed using a running average with a 5-ms window. This model
  shows a marginal and asymmetric explosion. The velocities are positive
  behind the shock, signaling an overall expanding flow, but the
  expansion rate is much smaller than for the n8.8 progenitors
  (Fig.~\ref{fig:velr.n8p8}).  Even after the explosion sets in, the
  velocity is still negative in regions behind the shock as a
  consequence of the partial fallback of the expanding plumes behind the
  shock.}
  \label{fig:summary.s9.0}
\end{figure*}

Some insight into the reason for the different evolutions can be gained
from the analysis of the accretion rate history of the progenitors, as
recently suggested by \citet{suwa:2014sqa} and \citet{muller:2016izw}.
Fig.~\ref{fig:mdot} shows the accretion rate at 500 km for all
progenitors in 1D, with the Baseline setup. Since the 1D progenitors
from \citet{sukhbold:2016a} fail to explode, these are ``intrinsic''
accretion rates, not affected by the explosion. The accretion rates for
the n8.8 and z9.6 are unaffected by the explosions up to the point where
they are shown (afterwards, they become negative as the inflow turns
into an outflow). It is easily seen that the accretion rates for
progenitors exploding in 1D decline steeply at very early times, which
sets them apart from ``normal'' massive stars, as also pointed out by
\citet{muller:2016izw}. We find that the accretion rate of the
10.0-$M_\odot$ progenitor is significantly higher than for the 9.0- and
11.0-$M_\odot$ models during the critical phase when the other two start
exploding. The sudden growth of the accretion rate of the 10.0-$M_\odot$
progenitor around ${\sim} 0.2\ {\rm s}$ after bounce is due to a small
density inversion present in the original progenitor profile
(Fig.~\ref{fig:progenitors}). Obviously, such a density inversion would
be Rayleigh-Taylor unstable and is not expected to be present in Nature.

Figures~\ref{fig:entropy2d.baseline} and \ref{fig:entropy2d.horowitz}
show snapshots of the entropy and of the pressure contrast, defined
following \citet{fernandez:2008ze} to be $r\,|\nabla p|/p$, for selected
progenitors at $0.15\ {\rm s}$, $0.2\ {\rm s}$, and $0.4\ {\rm s}$ after
bounce. In all our models, neutrino-driven convection is seeded in the
region behind the shock by perturbations induced by the dendritic grid
and develops rapidly following bounce (\ie, starting from ${\sim} 0.15\
{\rm s}$ after bounce). Large-scale shock-sloshing motions, possibly due
to the \ac{SASI} \citep{blondin:2002sm, foglizzo:2006fu}, are only
present at late times in models that fail to explode (e.g., the
10.0-$M_\odot$), after the shock has receded to less than $100\ {\rm
km}$ in radius. Due to the low post-bounce accretion rates, convection
is the dominant instability at early times and for all exploding models
\citep{foglizzo:2005xr, burrows:2012yk, murphy:2012id, mueller:2012ak,
ott:2012mr, couch:2013kma, fernandez:2013sqa, abdikamalov:2014oba}.

That said, convection appears to have a different role in the onset of
the explosion of the n8.8 compared to that of the other progenitors (for
the setups resulting in explosions).  For the n8.8, the shock starts
expanding rapidly already after about one convective overturn, before
the high-entropy plumes are able to reach it
(Figs.~\ref{fig:entropy2d.baseline} and \ref{fig:entropy2d.horowitz}).
The onset of this explosion is essentially spherically symmetric and
does not seem to be significantly aided by convection, although there
are small differences between 1D and 2D visible in
Fig.~\ref{fig:rshock}.

In the case of the \ac{ECSN}-like z9.6 and u8.1 progenitors, convective
plumes are able to reach the shock and appear to have an important role
in triggering the explosion, which is otherwise delayed or fails
altogether in 1D. The role of convective instabilities in the explosion
of the z9.6 progenitor was also studied by \citet{melson:2015tia}. Our
results are in qualitative agreement with theirs, but there are
quantitative differences. In particular, they found a significantly more
delayed explosion in 1D than in 2D and 3D, as a comparison of our 1D and
2D shock radius evolutions to theirs (Fig.~3 of
\citealt{melson:2015tia}) demonstrates.

Differently from the \ac{ECSN} and \ac{ECSN}-like progenitors, the 9.0-,
10.0-, 11.0-$M_\odot$ progenitors explode only after a few convective
overturns with the emergence of one or more large plumes that succeed in
pushing the shock to a sufficiently large radius to trigger a run-away
expansion. This behavior is commonly observed in 2D \ac{CCSN}
simulations \citep[\eg,][]{fernandez:2013sqa}.

After the explosion sets in, the shock and the material immediately
behind it expand almost self-similarly, with roughly constant
velocities. Behind them, we observe the emergence of a higher-entropy
neutrino-driven wind with dynamics similar to the one reported by
\citet{burrows:1987a} and \citet{burrows:1995ww}. For the n8.8, z9.6,
and u8.1 progenitors the wind is quasi-spherical, it produces weak
shocks visible in the pressure-contrast visualizations in
Figs.~\ref{fig:entropy2d.baseline} and \ref{fig:entropy2d.horowitz}, and
drives Rayleigh-Taylor instabilities as it pushes on the slower, heavier
material above. In the case of asymmetric explosions, the wind is
typically confined in ${\sim} 90^\circ$ wedges along the axis where it
drives the inflation of a large bubble, while fall-back accretion
continues along the equator. For these models, late-time accretion is
primarily responsible for the growth of the explosion energy, while for
the \ac{ECSN} and \ac{ECSN}-like explosions the explosion energy
injection is due to the wind. We caution the reader, however, that the
degree of asymmetry in the 9.0-, 10.0-, and 11.0-$M_\odot$ progenitor
explosions is likely to be artificially magnified by the assumption of
axisymmetry, and we speculate that the neutrino-driven wind will be
closer to spherical in full-3D simulations.

\begin{turnpage}
\begin{table*}
\caption{Summary of models and energy budget in the region $V: 100\ {\rm km}
\leq r \leq 20,\!000\ {\rm km}$. Values are given at final simulation
time.}
\label{tab:main}
\vspace{-1em}
\begin{center}
\begin{tabular}{lld{1.3}d{2.3}d{1.3}d{2.3}d{1.3}d{1.3}d{2.3}d{2.3}d{2.3}d{1.3}d{1.3}d{1.3}d{1.3}}
\toprule
Prog. &
Setup &
\myhead{$10^3 \xi_{2.5}$\tablenotemark{a}} &
\myhead{$E_{\rm bind}$\tablenotemark{b}} &
\myhead{$U$\tablenotemark{c}} &
\myhead{$E_0$\tablenotemark{d}} &
\myhead{$K_r$\tablenotemark{e}} &
\myhead{$K_\theta$\tablenotemark{f}} &
\myhead{$E_g$\tablenotemark{g}} &
\myhead{$E_{\rm tot}$\tablenotemark{h}} &
\myhead{$\dot{E}_{\rm tot}$\tablenotemark{i}} &
\myhead{$E_{\nu_e}$\tablenotemark{j}} &
\myhead{$E_{\bar{\nu}_e}$\tablenotemark{k}} &
\myhead{$E_{\nu_{\mu}}$\tablenotemark{l}} &
\myhead{$t_{\rm end}$\tablenotemark{m}} \\
&
&
&
\myhead{$[10^{50}\ {\rm erg}]$} &
\myhead{$[10^{50}\ {\rm erg}]$} &
\myhead{$[10^{50}\ {\rm erg}]$} &
\myhead{$[10^{50}\ {\rm erg}]$} &
\myhead{$[10^{50}\ {\rm erg}]$} &
\myhead{$[10^{50}\ {\rm erg}]$} &
\myhead{$[10^{50}\ {\rm erg}]$} &
\myhead{$[10^{50}\ {\rm erg}\ {\rm s}^{-1}]$} &
\myhead{$[10^{52}\ {\rm erg}]$} &
\myhead{$[10^{52}\ {\rm erg}]$} &
\myhead{$[10^{52}\ {\rm erg}]$} &
\myhead{$[{\rm s}]$} \\
\midrule
n8.8 & Baseline 1D & \myhead{-} & 0.000 & 0.328 & -0.058 & 0.994 & 0.000 & -0.061 & 1.202 & 0.038 & 1.346 & 0.830 & 2.582 &0.636 \\
n8.8 & ManyBody 1D & \myhead{-} & 0.000 & 0.354 & -0.062 & 1.158 & 0.000 & -0.064 & 1.386 & 0.058 & 1.383 & 0.871 & 2.977 &0.622 \\
n8.8 & Baseline 2D & \myhead{-} & 0.000 & 0.464 & -0.094 & 1.381 & 0.000 & -0.105 & 1.646 & 0.265 & 1.555 & 0.918 & 3.150 &0.619 \\
n8.8 & ManyBody 2D & \myhead{-} & 0.000 & 0.487 & -0.097 & 1.500 & 0.000 & -0.107 & 1.783 & 0.287 & 1.566 & 0.930 & 3.412 &0.602 \\
\midrule
u8.1 & Baseline 1D & 0.094 & -0.018 & 0.136 & -0.076 & 0.016 & 0.000 & -0.098 & -0.021 & 0.013 & 1.598 & 1.228 & 3.414 &0.888 \\
u8.1 & ManyBody 1D & 0.094 & -0.018 & 0.355 & -0.115 & 0.063 & 0.000 & -0.117 & 0.185 & 0.036 & 1.642 & 1.278 & 4.095 &0.914 \\
u8.1 & Baseline 2D & 0.094 & -0.018 & 0.645 & -0.204 & 0.523 & 0.001 & -0.105 & 0.861 & 0.085 & 2.037 & 1.526 & 5.145 &1.095 \\
u8.1 & ManyBody 2D & 0.094 & -0.018 & 0.715 & -0.211 & 0.636 & 0.001 & -0.112 & 1.029 & 0.113 & 2.039 & 1.523 & 5.557 &1.022 \\
\midrule
z9.6 & Baseline 1D & 0.076 & -0.008 & 0.173 & -0.052 & 0.076 & 0.000 & -0.055 & 0.143 & 0.019 & 1.518 & 1.157 & 3.374 &0.900 \\
z9.6 & ManyBody 1D & 0.076 & -0.008 & 0.244 & -0.073 & 0.256 & 0.000 & -0.046 & 0.381 & 0.021 & 1.593 & 1.235 & 4.072 &0.916 \\
z9.6 & Baseline 2D & 0.076 & -0.008 & 0.472 & -0.128 & 0.785 & 0.001 & -0.091 & 1.038 & 0.147 & 1.737 & 1.231 & 4.070 &0.771 \\
z9.6 & ManyBody 2D & 0.076 & -0.008 & 0.527 & -0.137 & 0.931 & 0.001 & -0.101 & 1.221 & 0.218 & 1.726 & 1.220 & 4.303 &0.722 \\
\midrule
9.0 & Baseline 1D & 0.038 & -0.021 & 0.401 & -0.206 & 0.034 & 0.000 & -0.269 & -0.040 & 0.012 & 1.652 & 1.298 & 3.287 &0.816 \\
9.0 & ManyBody 1D & 0.038 & -0.021 & 0.438 & -0.208 & 0.027 & 0.000 & -0.261 & -0.003 & 0.003 & 1.744 & 1.392 & 3.974 &0.848 \\
9.0 & Baseline 2D & 0.038 & -0.021 & 0.717 & -0.273 & 0.168 & 0.014 & -0.218 & 0.409 & 0.094 & 2.326 & 1.822 & 5.944 &1.454 \\
9.0 & Baseline Perturb 2D & 0.038 & -0.021 & 0.728 & -0.289 & 0.232 & 0.013 & -0.189 & 0.495 & 0.070 & 2.427 & 1.925 & 6.371 &1.645 \\
9.0 & ManyBody 2D & 0.038 & -0.021 & 0.841 & -0.292 & 0.295 & 0.012 & -0.229 & 0.627 & 0.202 & 2.361 & 1.846 & 6.655 &1.381 \\
\midrule
10.0 & Baseline 1D & 0.216 & -0.095 & 1.151 & -0.578 & 0.225 & 0.000 & -0.996 & -0.197 & 0.563 & 2.279 & 1.842 & 3.744 &0.807 \\
10.0 & ManyBody 1D & 0.216 & -0.095 & 1.226 & -0.560 & 0.184 & 0.000 & -0.904 & -0.053 & 0.477 & 2.458 & 2.016 & 4.627 &0.875 \\
10.0 & Baseline 2D & 0.216 & -0.095 & 0.788 & -0.394 & 0.054 & 0.000 & -0.370 & 0.078 & -0.012 & 3.764 & 3.142 & 8.968 &2.147 \\
10.0 & Baseline Perturb 2D & 0.216 & -0.095 & 0.996 & -0.463 & 0.086 & 0.000 & -0.592 & 0.027 & 0.186 & 3.059 & 2.450 & 6.505 &1.272 \\
10.0 & ManyBody Perturb 2D & 0.216 & -0.095 & 1.628 & -0.593 & 0.293 & 0.024 & -0.591 & 0.762 & 0.425 & 3.245 & 2.645 & 8.413 &1.575 \\
10.0 & ManyBody 2D & 0.216 & -0.095 & 1.126 & -0.482 & 0.087 & 0.000 & -0.615 & 0.117 & 0.303 & 3.082 & 2.463 & 7.183 &1.216 \\
\midrule
11.0 & Baseline 1D & 7.669 & -0.170 & 2.794 & -1.530 & 0.363 & 0.000 & -2.188 & -0.560 & 0.692 & 2.449 & 2.002 & 3.942 &0.924 \\
11.0 & ManyBody 1D & 7.669 & -0.170 & 3.019 & -1.536 & 0.339 & 0.000 & -2.174 & -0.351 & 0.781 & 2.558 & 2.105 & 4.724 &0.927 \\
11.0 & Baseline 2D & 7.669 & -0.170 & 3.751 & -1.573 & 0.700 & 0.068 & -1.628 & 1.318 & 0.497 & 3.136 & 2.539 & 7.033 &1.508 \\
11.0 & Baseline Perturb 2D & 7.669 & -0.170 & 3.585 & -1.544 & 0.379 & 0.068 & -1.816 & 0.671 & 0.619 & 3.311 & 2.707 & 7.428 &1.647 \\
11.0 & ManyBody 2D & 7.669 & -0.170 & 3.929 & -1.674 & 0.578 & 0.040 & -1.521 & 1.352 & 1.089 & 3.370 & 2.761 & 8.986 &1.780 \\
\bottomrule\vspace{-1em}
\end{tabular}
\tablenotetext{1}{Compactness parameter.}
\tablenotetext{2}{Envelope binding energy:
$\int_{r > 20,000\ {\rm km}} \rho\, \left[\epsilon - GM/r\right]\, \dd V$.}
\tablenotetext{3}{Total internal energy, excluding rest mass:
$\int_V \rho\, \epsilon\, \dd V$.}
\tablenotetext{4}{Total internal binding energy:
$\int_V \rho\, \epsilon_0\, \dd V$.}
\tablenotetext{5}{Radial kinetic energy:
$0.5\, \int_V \rho\, v_r^2\, \dd V$.}
\tablenotetext{6}{Non-radial kinetic energy:
$0.5\,\int_V \rho\, (r\, v_\theta)^2\, \dd V$.}
\tablenotetext{7}{Gravitational energy:
$-\int_V \rho\, GM/r\, \dd V$.}
\tablenotetext{8}{Total energy in the region of interest: internal, binding,
kinetic, and gravitational.}
\tablenotetext{9}{Rate of change of the net energy at the end of the simulation
estimated from the last 10 ms of data.}
\tablenotetext{10}{Total energy radiated in $\nu_e$.}
\tablenotetext{11}{Total energy radiated in $\bar{\nu}_e$}
\tablenotetext{12}{Total energy radiated in ``$\nu_{\mu}$''.}
\tablenotetext{13}{Final simulation time (in seconds after bounce). This is the
post-bounce time the maximum shock radius exceeds 19000 km for successful explosions.}
\end{center}
\end{table*}
\end{turnpage}

As a representative example of an \ac{ECSN}-like explosion, we show in
Figs.~\ref{fig:summary.n8p8} and \ref{fig:velr.n8p8} a summary of the
evolution of the n8.8 progenitor in 1D and 2D, evolved with the Baseline
setup. Despite their relatively similar average shock trajectories and
the fact that the n8.8 shock remains nearly spherical in both 1D and 2D
simulations, there
are substantial differences between the 1D and 2D explosions. In
particular, the entropy in the 1D model exceeds that of the 2D model,
while the velocity of the neutrino-driven wind in the 2D model exceeds
that in the 1D model, roughly by a factor of two. This is partly due to
the tendency of 1D models to create low-density regions that become
overheated by neutrinos. More importantly, starting from ${\sim} 0.2\
{\rm s}$ after bounce, the 2D simulation shows significantly larger
neutrino luminosities (see Sec.~\ref{sec:pns} for a detailed account).
This results in stronger neutrino-driven winds, with increased
velocities (see Fig.~\ref{fig:velr.n8p8}) and correspondingly smaller
expansion timescales.

The 9.0-, 10.0-, and 11.0-$M_\odot$ progenitors evolve in a
qualitatively different way.  Fig.~\ref{fig:summary.s9.0} shows the
evolution in 2D of the 9.0-$M_\odot$ progenitor with the Baseline setup, which we
take as representative of regular (successful) \acp{CCSN} from a
low-mass progenitor. The explosion of the 9.0 follows ${\sim} 100$ ms of
shock stagnation and is triggered at the time when the Si/O interface is
accreted. As already mentioned, the explosion is asymmetric. It is also
marginal, with small velocities immediately below the shock and a degree
of sustained fallback at the end of the simulation. The z9.6 and u8.1
progenitors' dynamics are similar to that of the regular \ac{CCSN}
progenitors at early times, during the shock stagnation epochs. However,
their explosion is also triggered by the sharp accretion rate drop in an
almost spherical way, and then evolves in a qualitatively similar way to
that of the n8.8 progenitor.

\section{Explosion Energetics}
\label{sec:explene}

\begin{figure*}
  \begin{minipage}{\columnwidth}
    \includegraphics[width=\columnwidth]{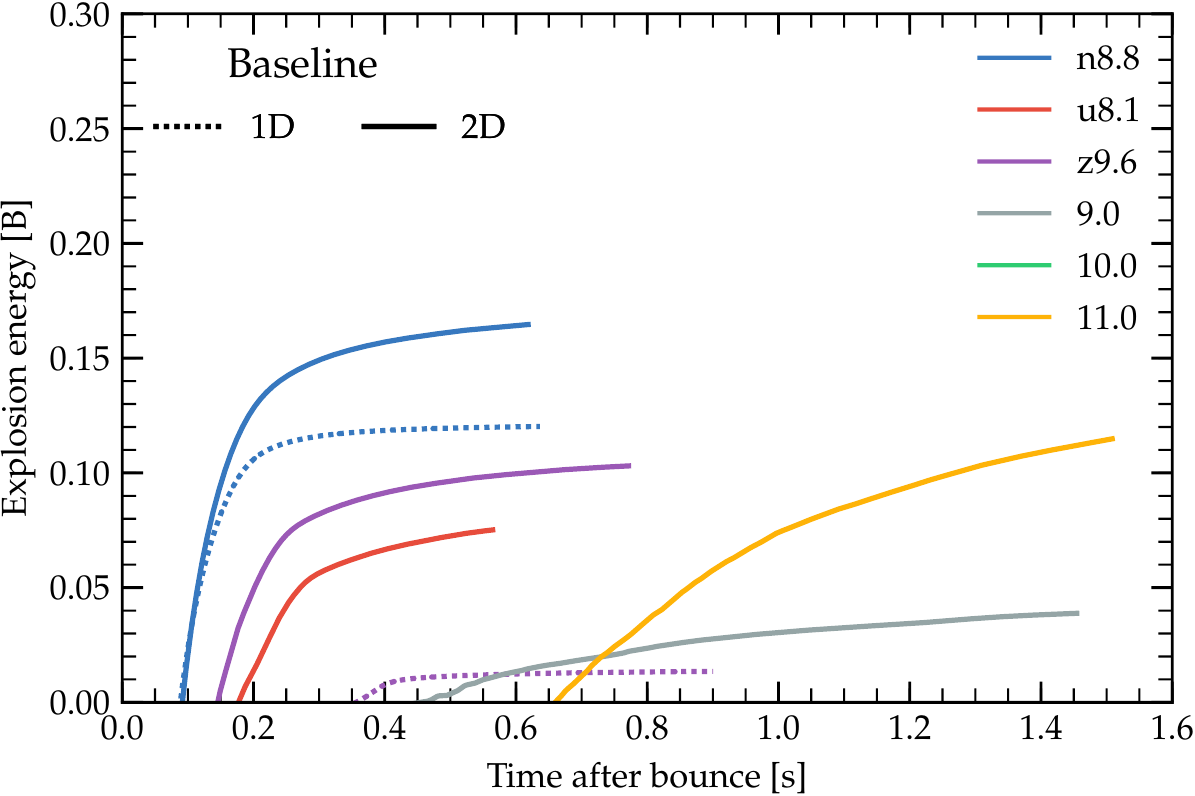}
  \end{minipage}
  \hfill
  \begin{minipage}{\columnwidth}
    \includegraphics[width=\columnwidth]{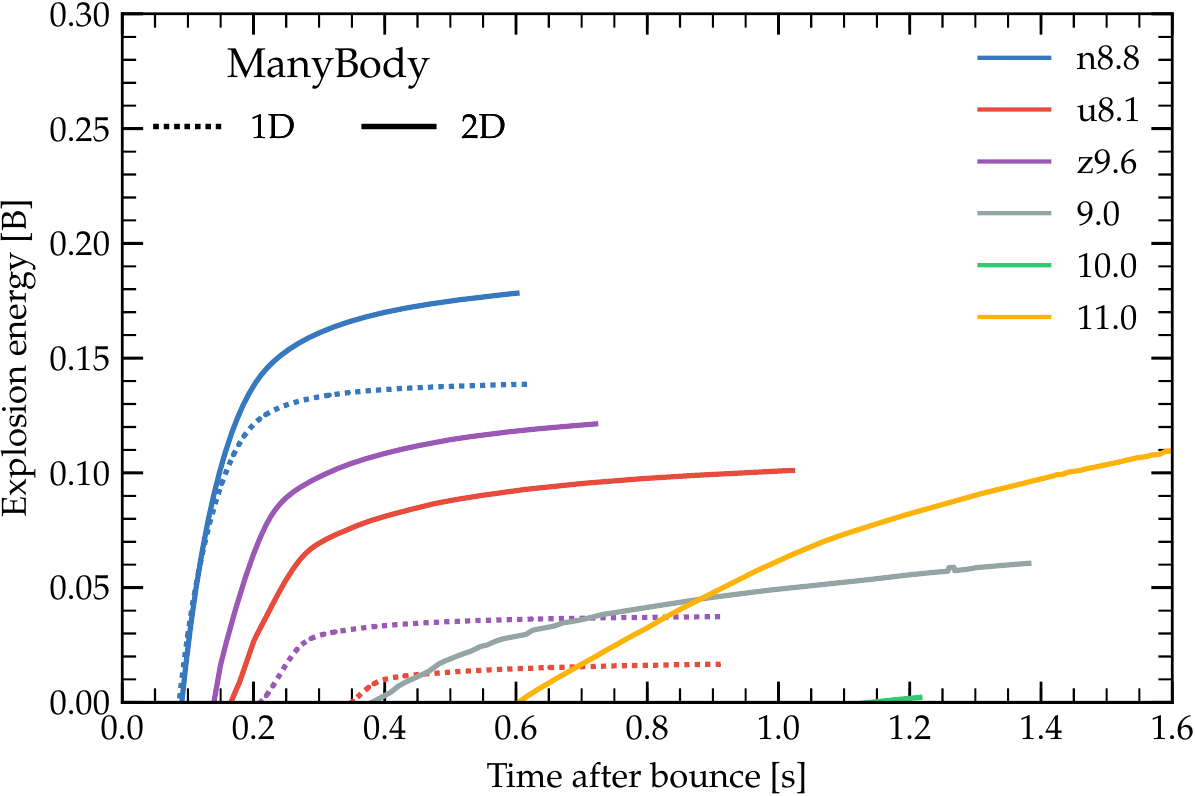}
  \end{minipage}
  \caption{Net explosion energy in Bethe ($10^{51}\ {\rm erg}$) in the
  region $100\ {\rm km} \leq r \leq 20,\!000\ {\rm km}$ for all
  progenitors in 1D and 2D with our Baseline setup (\emph{left panel})
  and with the inclusion of many-body corrections (\emph{right panel}).
  We account for the binding energy of the envelope in the estimate of
  the explosion energy. The curves are smoothed using a running average
  with a 5-ms window. Many-body corrections have a ${\sim}10\%$ level
  impact on the explosion energies of all models, including those that
  explode in 1D.}
  \label{fig:explene}
\end{figure*}

We estimate explosion energies using a fixed-volume energy analysis
similar to that in \citet{bruenn:2014qea}. We consider the region $r
\geq 100\ {\rm km}$, and we compute the integrated total energy $E_{\rm
tot}$ as the sum of internal $U$, kinetic $K$, and gravitational binding
energy $E_g$. To this, we subtract the zero-temperature energy of the
material $E_0$ computed from the \ac{EOS} used for the evolution. We
remark that $E_{\rm tot}$ includes the contribution of the binding
energy of the material exterior to the shock but interior to the
computational domain. The values of these quantities at the end of our
simulations are given in Tab.~\ref{tab:main}. There, we also quote the
total energy liberated in each species of neutrinos $E_{\nu_e}$,
$E_{\bar{\nu}_e}$, and $E_{\nu_\mu}$. The final explosion energy can be
estimated by subtracting in absolute value the binding energy of the
material exterior to $20,\!000\ {\rm km}$ from $E_{\rm tot}$. This is
shown, as a function of time, in Fig.~\ref{fig:explene}. Note that, at
late times, the net energy in the integration region can take (small)
positive values also for failing models, \eg, for the 10.0-$M_\odot$
progenitor. This is mostly because our metric includes with a positive
sign also the kinetic energy of infalling fluid elements. Furthermore,
we remark that, because of the \ac{GR} corrections included in our
treatment of gravity, the total energy is not conserved. Instead, the
gravitational potential decreases by several percent over the duration
of our simulations, because of neutrino losses.

We find explosion energies ranging from a few percent of a Bethe ($1\
{\rm B} \equiv 10^{51}\ {\rm erg}$), like the 9.0-Baseline 2D run, to
values in excess of $0.17$ Bethes, for the n8.8-ManyBody 2D simulation.
The explosion energy we estimate for the z9.6 progenitor with the
Baseline setup is ${\sim} 50\%$ larger than that reported by
\citet{melson:2015tia} for their 2D \textsc{Prometheus-Vertex}
simulation. The estimated explosion energies for the z9.6 and u8.1 in 2D
are also similarly larger than those of \textsc{Coconut-Vertex} in 2D,
as quoted in \citet{wanajo:2017cyq}. The discrepancy is somewhat larger
for the n8.8 progenitor, where we estimate an explosion energy almost a
factor two larger than reported by \citet{wanajo:2017cyq}. On the other
hand, the explosion energies for the z9.6 and n8.8 in 1D ($0.01\
{\rm B}$ and $0.12\ {\rm B}$) are in good agreement with those reported
by the Garching group \citep{melson:2015tia, kitaura:2005bt},
suggesting that the discrepancies might be due to multi-dimensional
effects. Note, however, that in \citet{janka:2007di} the
Garching group reported a $20\%$ smaller explosion energy for the n8.8
progenitor compared to our results and their own previous calculations
\citep{kitaura:2005bt}. Explosion energies for the progenitors from
\citet{sukhbold:2016a} have not been reported before, so no comparison
is possible.

The reduction of neutral current interactions in the ManyBody setup
yields an increase in the explosion energies of 10\% to 50\%, depending
on the model. The amplification is particularly large for the
9.0-$M_\odot$ progenitor in 2D and the z9.6 progenitor in 1D, where the
ManyBody setup boosts the explosion energy by ${\sim} 50\%$. These large
amplifications are due to the proximity of these progenitors to
criticality, which amplifies their sensitivity to relatively small
changes in the input microphysics.

The role of perturbations on the explosion energy (Tab.~\ref{tab:main})
is not completely clear. In the case of the 9.0- and 10.0-$M_\odot$
progenitors, the inclusion of perturbations is beneficial. The first
explodes with slightly larger (${\sim} 10\%$) explosion energy than
without perturbations. The second goes from a failed to a successful,
albeit underenergetic, explosion with the introduction of perturbations
in combination with many body corrections to neutral current
interactions. Somewhat surprisingly, in the case of the 11.0-$M_\odot$
progenitor, the inclusion of perturbations results in a reduction of the
explosion energy by a factor 2. The reason is that 11.0-Perturb
explosion entrains more bound mass and the ejecta lose more energy,
while doing work on the infalling envelope of the star. On the other
hand, note that the explosion energies for the 11.0-$M_\odot$ progenitor
are still growing significantly at the end of our simulations, so the
difference between the 11.0-Perturb and 11.0-Baseline models might be
only transitory.

It is important to keep in mind that the explosion energies we quote are
not final. Indeed the estimated explosion energy is still growing
significantly at the time we stop the calculation for many of our
simulations. This is not surprising in the light of the results of
\citet{muller:2015dia}, who studied the development of an explosion in
the 11.2-$M_\odot$ progenitor from \citet{woosley:2002zz} and found the
explosion energy to saturate only after several seconds. On the other
hand, the explosion energies saturate very rapidly for the n8.8, u8.1,
and z9.6 progenitors and appear to have converged within the simulation
time.

\begin{figure}
  \includegraphics[width=\columnwidth]{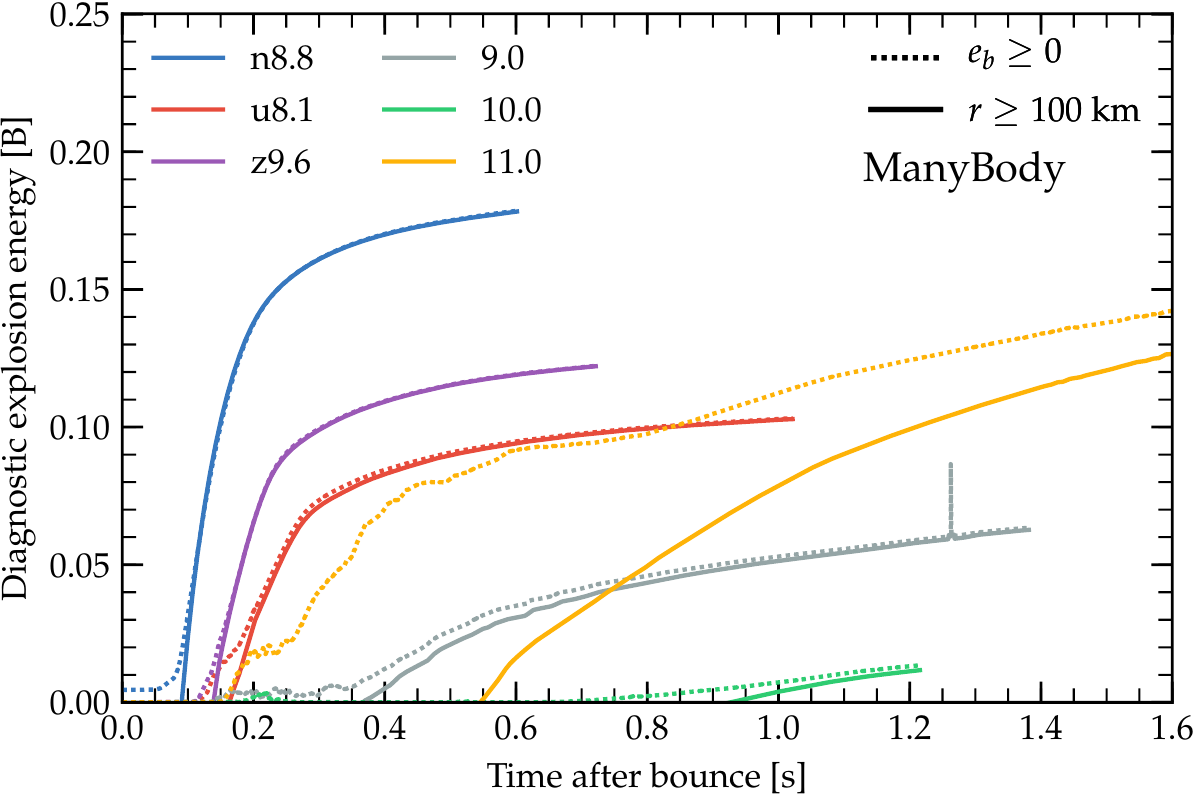}
  \caption{Diagnostic explosion energy $E_{\rm tot}$ in Bethe ($\equiv
  10^{51}\ {\rm erg}$) integrated over the entire region where $r \geq
  100\ {\rm km}$, or only for elements with positive net energy $e_{b}
  \geq 0$. The binding energy of the envelope has not been included in
  either calculation.}
  \label{fig:explene.methods}
\end{figure}

Another caveat is that our estimate of the explosion energy is more
conservative than the commonly used ``diagnostic explosion energy.'' The
former is computed as $E_{\rm tot}$, but only integrated over unbound
and/or radially expanding fluid elements \citep[\eg,][]{buras:2005tb,
mueller:2012is} and does not include the overburden of the material
exterior to the shock. A comparison between the two is given in
Fig.~\ref{fig:explene.methods}. There, we compute the diagnostic energy
as the integral of the total energy density $e$ minus the zero-point
energy $e_0$, $e_{\rm tot} = e - e_0$, either integrated over regions
where $e_{\rm tot} \geq 0$, or over $100\ {\rm km} \leq r \leq 20,\!000\
{\rm km}$. For clarity, we did not include the binding energy of the
envelope when computing $E_{\rm tot}$ in this plot, since its inclusion
in the diagnostic explosion energy would be inconsistent.  Beside this
difference, the $r \geq 100\ {\rm km}$ diagnostic energies in
Fig.~\ref{fig:explene.methods} are identical to the estimated explosion
energies in Fig.~\ref{fig:explene}. Obviously, the diagnostic energy
integrated only over $e_{\rm tot} \geq 0$ or $r \geq 100\ {\rm km}$
should converge to the same value after a sufficiently long time.  This
is indeed the case for most of our models, especially the
\ac{ECSN}/\ac{ECSN}-like explosions, where the explosion is close to
spherically symmetric. However, significant differences persist until
the end of our simulations for some progenitors. For example, in the
11.0-ManyBody run, the shock starts expanding ${\sim} 200\ {\rm ms}$
after bounce, and some material becomes unbound. However, the energy
behind the shock becomes sufficient to overcome the overburden only at
later times, when the \ac{PNS}\acused{NS} wind becomes violent enough to
create high-entropy bubbles behind the shock and the initially bound
material at $r\geq 100\ {\rm km}$ has accreted or has become unbound.

\section{Neutrino Radiation}
\label{sec:nurad}
\begin{figure*}
  \begin{minipage}{\columnwidth}
    \includegraphics[width=\columnwidth]{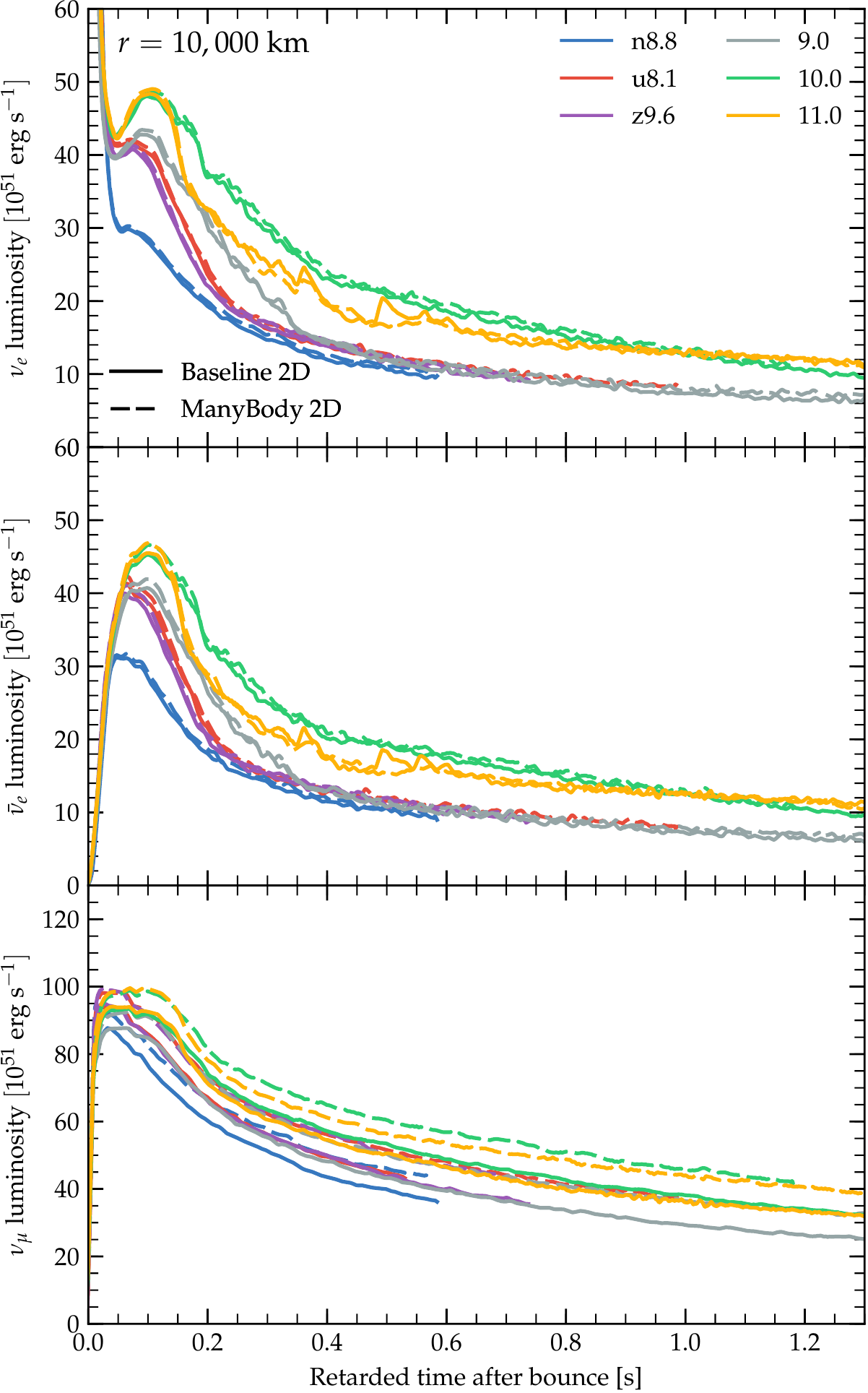}
  \end{minipage}
  \hfill
  \begin{minipage}{\columnwidth}
    \includegraphics[width=\columnwidth]{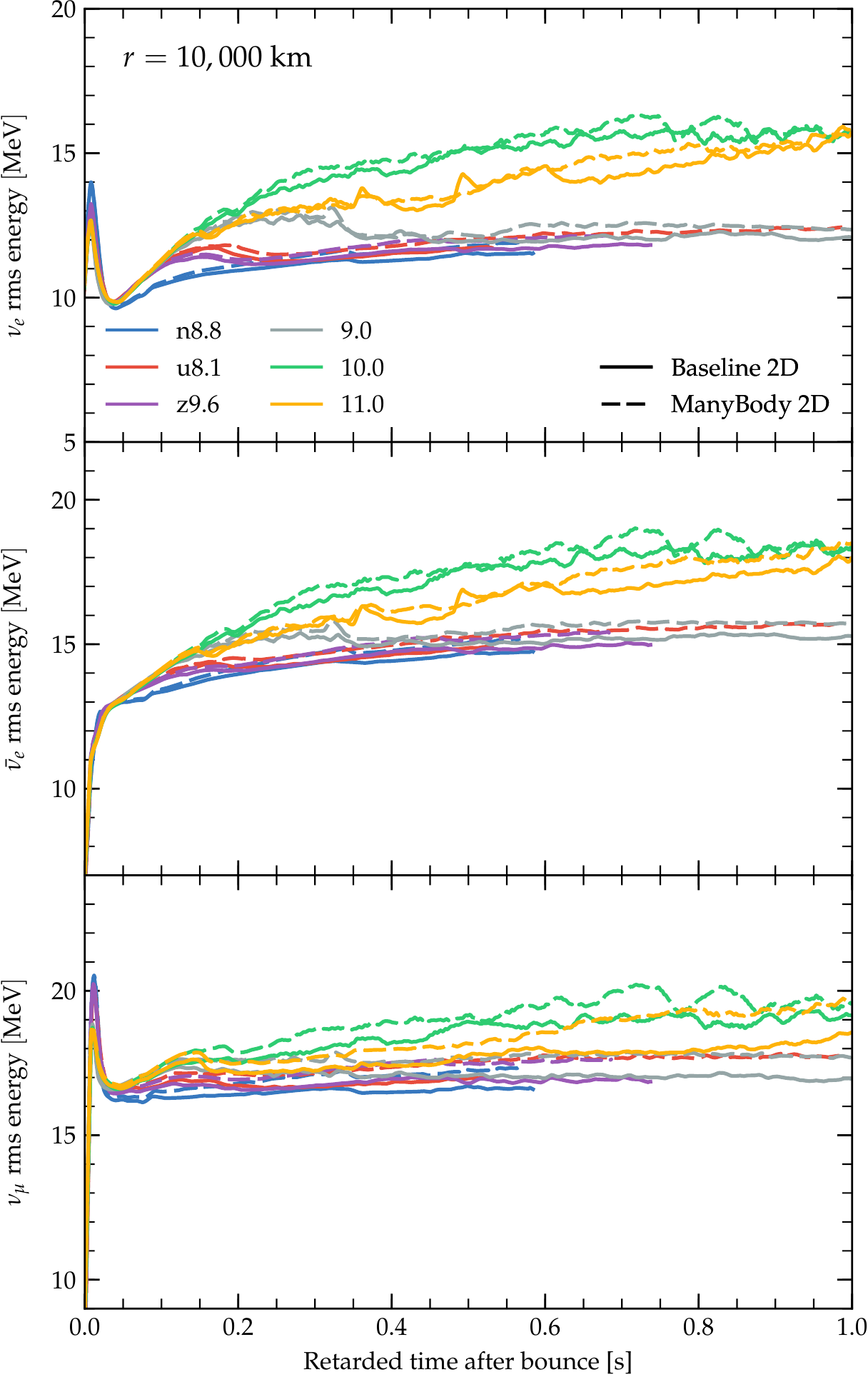}
  \end{minipage}
  \caption{Neutrino luminosity (\emph{left panel}) and rms energies
  (\emph{right panel}) at $10,\!000$ km as a function of the retarded
  time for all progenitors. Here, $\nu_\mu$ denotes the sum of all
  heavy-lepton neutrino species. The curves are smoothed using a running
  average with a 5-ms window. The many-body corrections implemented in
  our ManyBody setup result in a several percent increase in the
  heavy-lepton neutrino luminosity and a slight increase in the average
  energies for all neutrino species.}
  \label{fig:luminosity}
\end{figure*}

\begin{figure*}
  \begin{minipage}{\columnwidth}
    \includegraphics[width=\columnwidth]{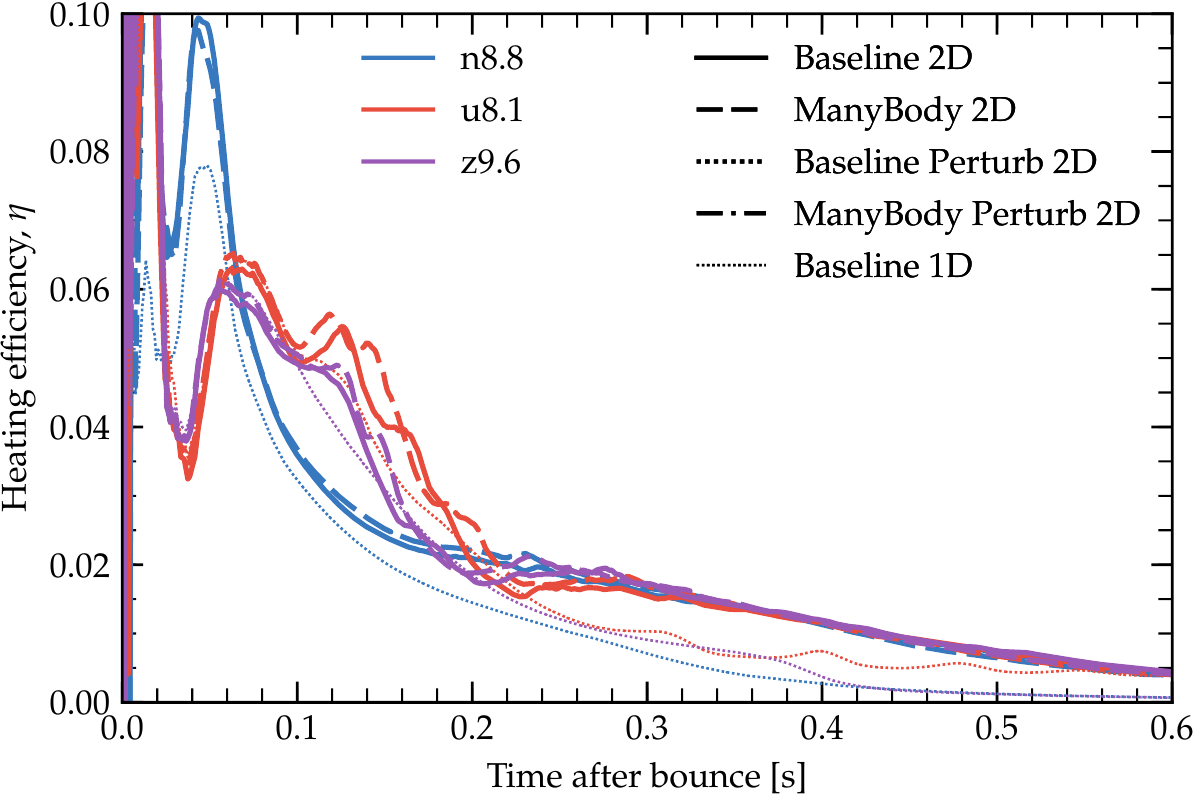}
  \end{minipage}
  \hfill
  \begin{minipage}{\columnwidth}
    \includegraphics[width=\columnwidth]{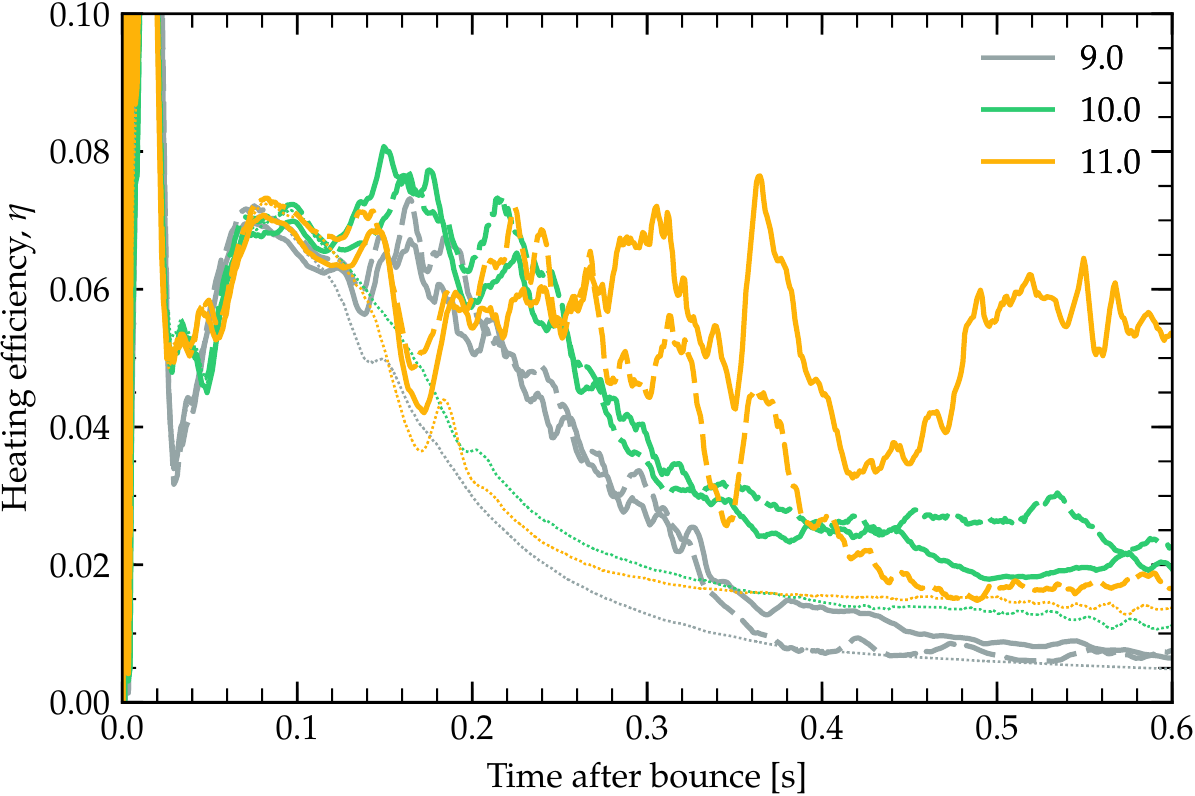}
  \end{minipage}
  \caption{Heating efficiency, $\eta$, for the n8.8, u8.1, and z9.6
  (\emph{left panel}), and 9.0-, 10.0-, and 11.0-$M_\odot$ progenitors
  (\emph{right panel}). The curves are smoothed using a running average
  with a 5-ms window. Many-body effects result in a hardening of the
  neutrino radiation which, in turn, leads to a slightly better coupling
  of $\nu_e$ and $\bar{\nu}_e$ with the material in the gain region. One
  exception is the 11.0-$M_\odot$ progenitor in which the earlier shock
  expansion induced by the many-body effects leads to a decrease of the
  accretion rate and, consequently, of the mass in the gain region. This
  in turn results in lower heating efficiency in the ManyBody case.}
  \label{fig:heateff}
\end{figure*}

We extract the properties of the neutrino radiation on a sphere placed
at $10,\!000\ {\rm km}$ from the center. Angle-averaged neutrino
luminosities and rms neutrino energies are shown in
Fig.~\ref{fig:luminosity}.  These are shown in the lab frame at
infinity.  Note that \fornax evolves the fluid-frame neutrino-radiation
moments and does not output the Eddington factor used for the evolution.
For this reason, we convert the code output to the lab frame under the
simplifying assumption of a forward-peaked radial neutrino distribution
function. This assumption is not valid at the time the shock crosses
10,000 km and results in small jumps which are particularly evident in
the rms energies of heavy-lepton neutrinos for the n8.8 progenitor.

We find the ManyBody setup to result in slightly higher neutrino
luminosities and average energies. The heavy-lepton neutrino
luminosities are the most clearly affected and increase by ${\sim}
10\%$, since their opacity is dominated by neutrino-nucleon scattering.
However, the average energies for all neutrino species increase because
of the accelerated contraction rate of the \ac{PNS} with the ManyBody
setup (see Sec.~\ref{sec:pns}).


We quantify the degree of coupling between the neutrino radiation and
the accretion flow in terms of the heating efficiency parameter $\eta$,
defined as the ratio between the heating rate by neutrinos in the gain
region, \ie, the region bounded by the \ac{PNS} and the shock with
positive net neutrino heating, and the sum of the $\nu_e$ and
$\bar{\nu}_e$ luminosities at infinity \citep[\eg,][]{marek:2007gr,
mueller:2012is, mueller:2012ak}. We show this quantity as a function of
time in Fig.~\ref{fig:heateff} for 1D and 2D models with the Baseline
and the ManyBody setups.  We recall that the same analysis was performed
by \citet{mueller:2012ak} for the u8.1 progenitor. We find good
agreement with their heating efficiency and neutrino luminosities.
For the other models, the overall trend in $\eta$ is that progenitors
with larger accretion rates also show larger heating efficiencies.

Spherically-symmetric (1D) simulations have significantly smaller
heating efficiencies. The increased heating efficiency in 2D is in part
due to the longer dwelling time of material in the gain region
\citep{burrows:1995ww, murphy:2008dw, dolence:2012kh} or, equivalently
\citep{mueller:2012is} to the growth of the mass in the gain region.

The many-body corrections implemented in the ManyBody setup result in a
slight increase of the heating efficiency. At least at early times,
before the evolutionary paths of the Baseline and ManyBody simulations
start to diverge, this improvement can be attributed to the hardening of
the neutrino spectra with the ManyBody setup, which results in a more
tight coupling with the material. After the explosions set in, the
differences between the Baseline and ManyBody efficiencies are in good
part due to the fact that the ManyBody explosions are more spherical and
entrain more mass.

\section{Protoneutron Stars}
\label{sec:pns}
\begin{figure}
  \includegraphics[width=\columnwidth]{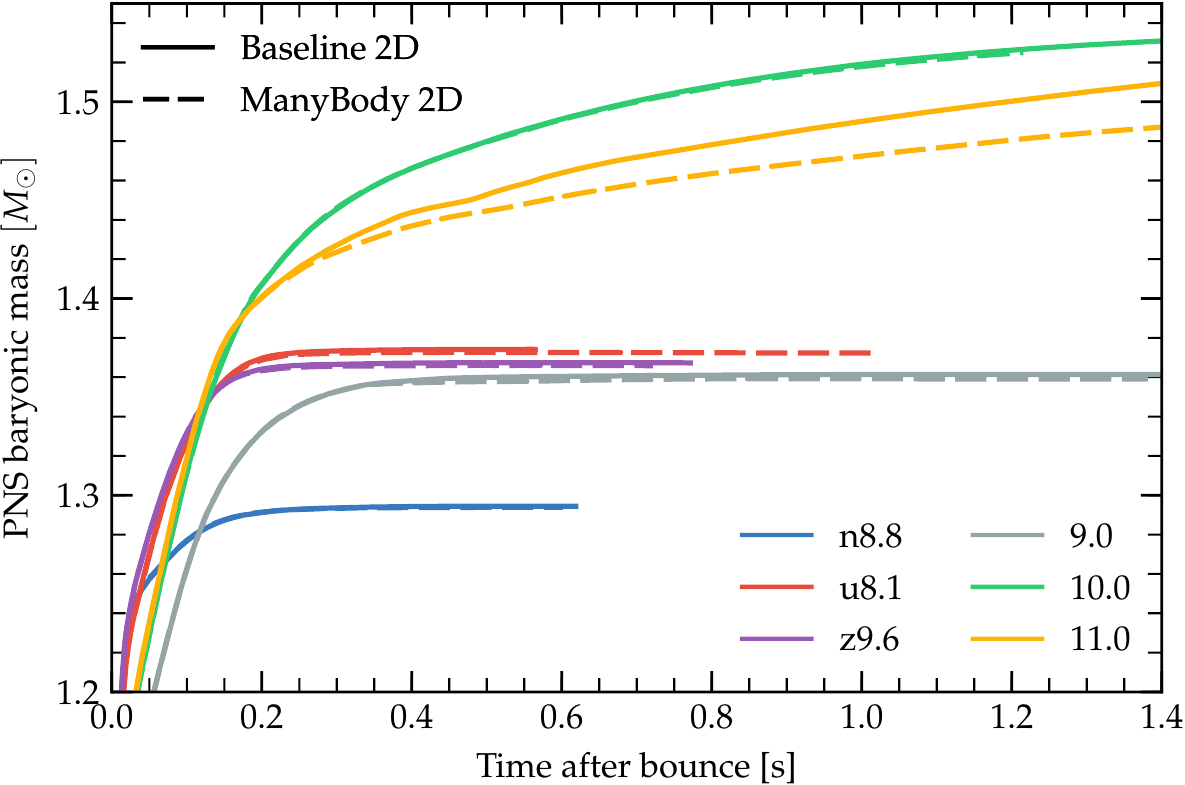}
  \caption{PNS baryonic masses for the 2D simulations with the Baseline
  and ManyBody setups. The curves are smoothed using a running average
  with a 5-ms window. Many-body corrections to the neutrino-nucleon
  scattering opacities result in earlier, more vigorous explosions and,
  consequently, slightly smaller \ac{PNS} masses.}
  \label{fig:pnsmass}
\end{figure}

\begin{figure*}
  \begin{minipage}{\columnwidth}
    \includegraphics[width=\columnwidth]{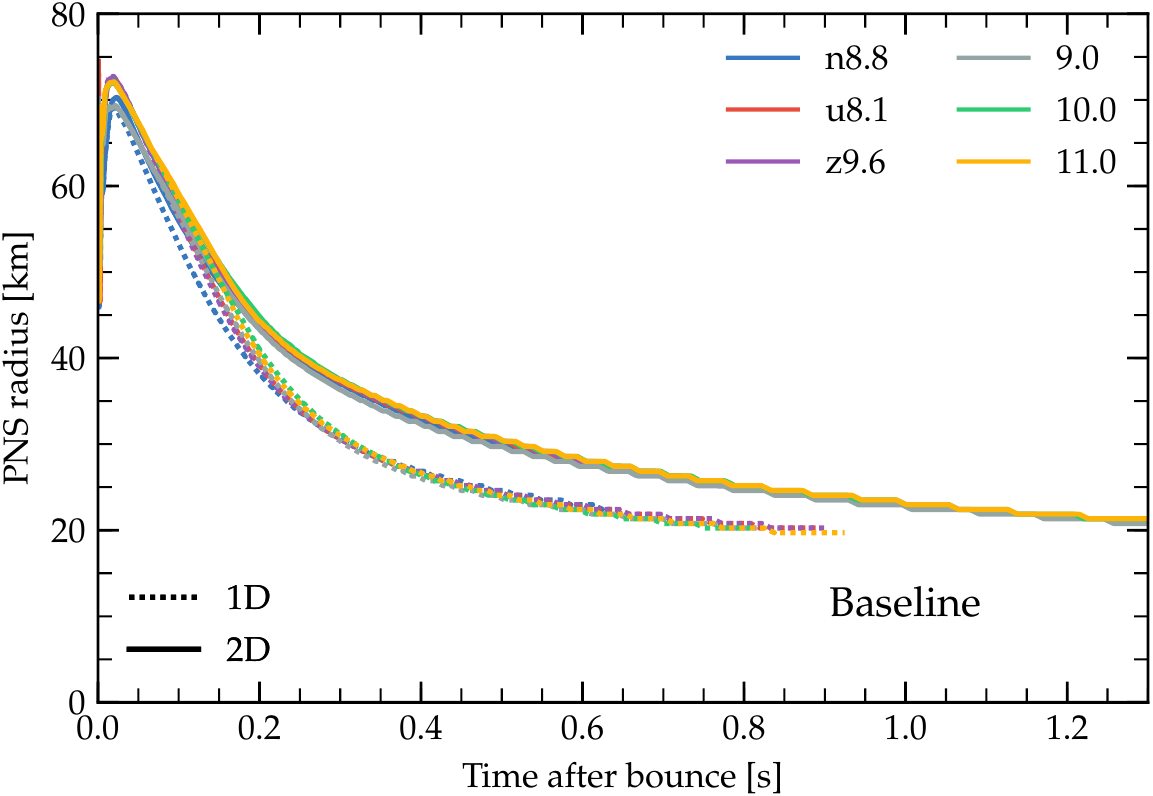}
  \end{minipage}
  \hfill
  \begin{minipage}{\columnwidth}
    \includegraphics[width=\columnwidth]{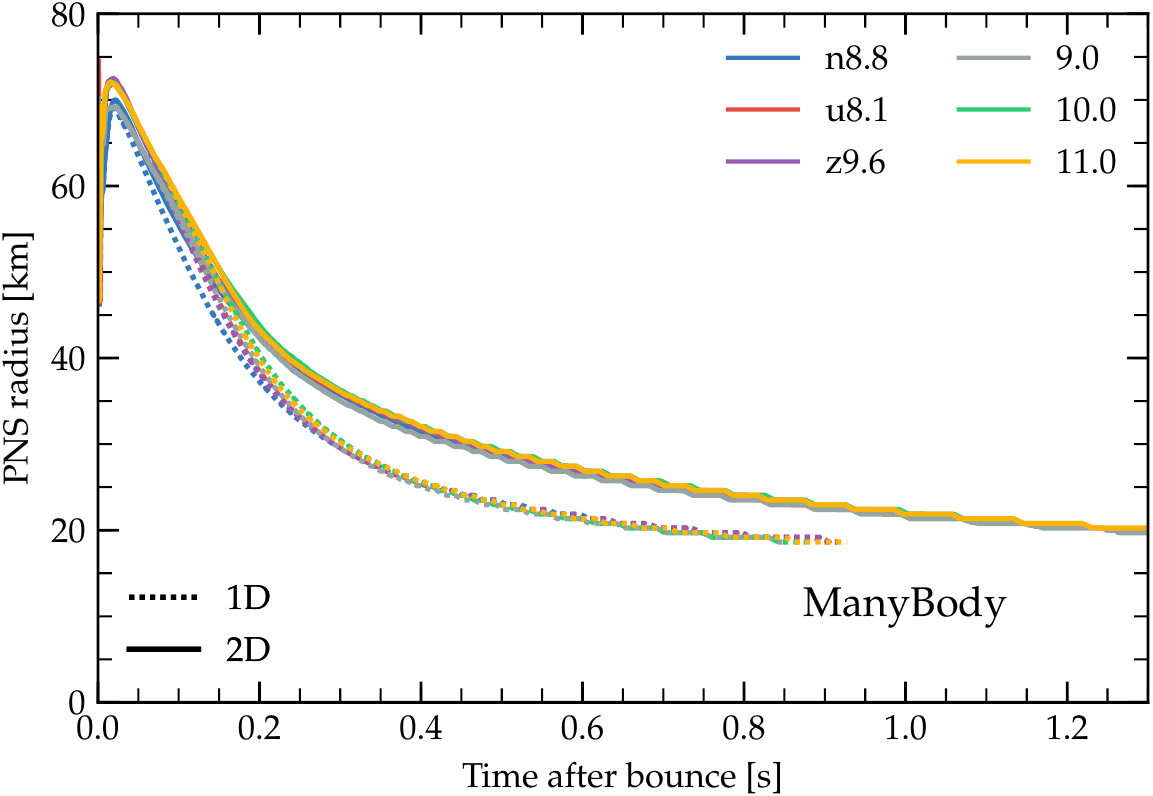}
  \end{minipage}
  \caption{PNS radii in 1D and 2D with Baseline physics (\emph{left
  panel}) and with many-body corrections (\emph{right panel}). Curves
  are smoothed using a running average with a 5-ms window. Many-body
  corrections result in slightly faster PNS contraction rates, however
  the largest differences are between 1D and 2D simulations.
  One-dimensional models predict faster contraction rates of the PNS
  starting from ${\sim} 0.2\ {\rm s}$ after bounce.}
  \label{fig:pnsradius}
\end{figure*}

As is commonly done in the \ac{CCSN}-mechanism literature, we define as
\ac{PNS} radius the radius at which the angle-averaged density is
$10^{11}\ {\rm g}\ {\rm cm}^{-3}$. We monitor \ac{PNS} radii and the
baryonic mass they enclose to estimate the final remnant radius. These
quantities are shown in Figs.~\ref{fig:pnsmass} and \ref{fig:pnsradius}.
\ac{PNS} masses and accretion rates at the end of our simulations are
also given in Tab.~\ref{tab:pns}. There we also quote the corresponding
gravitational mass for a cold, deleptonized \ac{NS} estimated using the
approximate fit of \citet{timmes:1995kp}.

The \ac{PNS} masses, as was the case for the explosion energy, are still
not converged for the 11.0-$M_\odot$ progenitor at the end of our
simulations. The \ac{PNS} mass for the 10.0-$M_\odot$ model is obviously
converged only with the ManyBody setup with perturbations, which
explodes, while black-hole formation would appear inevitable for the
other setups.

Notwithstanding these caveats, we find that all our progenitors produce
\acp{PNS} with gravitational masses below $1.4\ M_\odot$. The n8.8
progenitor produces \acp{PNS} with gravitational masses as low as
$1.188\ M_\odot$, suggesting that an \ac{ECSN}/\ac{ECSN}-like explosion
might be able to explain the origin of the low-mass companion in the
double \ac{NS} system J0453+1559. This has been recently measured, using
the advance of periastron and the Shapiro delay, to have a mass of
$1.174 \pm 0.004\ M_\odot$ \citep{martinez:2015mya}. This scenario is
also plausible in light of the study by \citet{tauris:2015xra}, who
showed that an \ac{ECSN} is a possible outcome of the evolution of
ultra-stripped metal cores in tight binaries, such as those producing
relativistic double-\ac{NS} systems like J0453+1559.

We caution the reader that previous studies reported somewhat
larger \ac{PNS} masses for the n8.8 progenitor. The Garching group
reported a final (baryonic) \ac{PNS} mass of $1.366\ M_\odot$
\citep{hudepohl:2010a}, while \citet{fischer:2009af} reported a final
\ac{PNS} mass of $1.347\ M_\odot$. The origin for our smaller \ac{PNS}
masses is probably due to our neglecting of electron capture on heavy
nuclei and nuclear burning during infall for this model. Both could
slightly increase the \ac{PNS} mass. As argued by \citet{burrows:1985a},
nuclear burning in the supersonically infalling material will accelerate
the collapse. Electron capture on heavy nuclei will decrease the
pressure support in the core and further accelerate the collapse.

\begin{table}
\caption{PNS star masses at the final simulation time.}
\label{tab:pns}
\vspace{-1em}
\begin{center}
\begin{tabular}{lld{1.3}d{1.3}d{1.3}}
\toprule
Prog. &
Setup &
\myhead{$M_{{\rm Baryon}}$\tablenotemark{a}} &
\myhead{$M_{{\rm Grav}}$\tablenotemark{b}} &
\myhead{$\dot{M}_{{\rm Baryon}}$\tablenotemark{c}} \\
&
&
\myhead{$[M_{\odot}]$} &
\myhead{$[M_{\odot}]$} &
\myhead{$[M_{\odot}\ {\rm s}^{-1}]$} \\
\midrule
n8.8 & Baseline 1D & 1.300 & 1.193 & 0.003 \\
n8.8 & ManyBody 1D & 1.299 & 1.193 & 0.003 \\
n8.8 & Baseline 2D & 1.294 & 1.188 & 0.004 \\
n8.8 & ManyBody 2D & 1.294 & 1.188 & 0.005 \\
\midrule
u8.1 & Baseline 1D & 1.391 & 1.270 & 0.005 \\
u8.1 & ManyBody 1D & 1.385 & 1.265 & 0.000 \\
u8.1 & Baseline 2D & 1.374 & 1.256 & 0.008 \\
u8.1 & ManyBody 2D & 1.372 & 1.254 & 0.000 \\
\midrule
z9.6 & Baseline 1D & 1.379 & 1.260 & 0.000 \\
z9.6 & ManyBody 1D & 1.376 & 1.257 & 0.000 \\
z9.6 & Baseline 2D & 1.367 & 1.250 & 0.002 \\
z9.6 & ManyBody 2D & 1.366 & 1.249 & 0.003 \\
\midrule
9.0 & Baseline 1D & 1.369 & 1.252 & 0.013 \\
9.0 & ManyBody 1D & 1.369 & 1.252 & 0.011 \\
9.0 & Baseline 2D & 1.361 & 1.245 & 0.000 \\
9.0 & Baseline Perturb 2D & 1.359 & 1.243 & 0.001 \\
9.0 & ManyBody 2D & 1.359 & 1.243 & 0.001 \\
\midrule
10.0 & Baseline 1D & 1.509 & 1.368 & 0.067 \\
10.0 & ManyBody 1D & 1.512 & 1.371 & 0.056 \\
10.0 & Baseline 2D & 1.540 & 1.394 & 0.008 \\
10.0 & Baseline Perturb 2D & 1.528 & 1.385 & 0.024 \\
10.0 & ManyBody Perturb 2D & 1.510 & 1.369 & 0.003 \\
10.0 & ManyBody 2D & 1.525 & 1.382 & 0.026 \\
\midrule
11.0 & Baseline 1D & 1.515 & 1.373 & 0.093 \\
11.0 & ManyBody 1D & 1.513 & 1.372 & 0.089 \\
11.0 & Baseline 2D & 1.514 & 1.372 & 0.034 \\
11.0 & Baseline Perturb 2D & 1.519 & 1.377 & 0.024 \\
11.0 & ManyBody 2D & 1.496 & 1.358 & 0.024 \\
\bottomrule\vspace{-1em}
\end{tabular}
\tablenotetext{1}{PNS baryonic mass.}
\tablenotetext{2}{PNS gravitational mass.}
\tablenotetext{3}{PNS accretion rate.}
\end{center}
\end{table}

We find the \ac{PNS} radii (Fig.~\ref{fig:pnsradius}) to follow tracks
that are largely independent of the progenitor or the \ac{PNS} mass, as
do \citet{bruenn:2014qea} and \citet{summa:2015nyk}. The reason is that
the density drops sharply at the surface of the \ac{PNS} so that the
ambient pressure has a negligible influence on the structure of the
central object. This is determined by the competition between its
internal pressure and gravity. Instead, the radii are sensitive to
changes in the microphysical treatment, which determines the rate at
which the \ac{PNS} deleptonizes and loses thermal support, and to the
dimensionality (1D vs.~2D). The impact of the microphysics is easily
understood from the fact that the contraction of the \ac{PNS} is mostly
set by the rate of deleptonization and core cooling. These in turn
depend in the first second after bounce on the neutrino opacity of
matter at densities between $10^{11}$ and $10^{13}\ {\rm g}\ {\rm
cm}^{-3}$. For instance, the many-body corrections included in the
ManyBody setup result in a faster deleptonization and contraction of the
\ac{PNS}.

\begin{figure*}
  \begin{minipage}{\columnwidth}
    \includegraphics[width=\columnwidth]{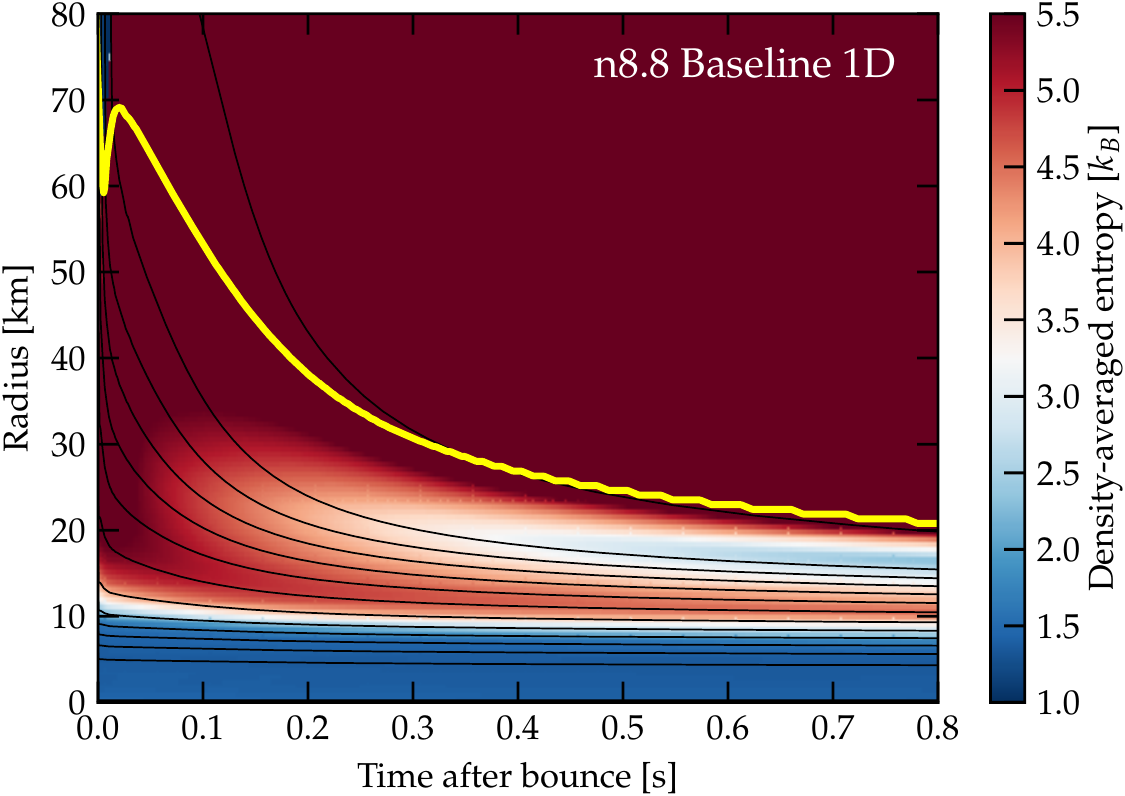}
  \end{minipage}
  \hfill
  \begin{minipage}{\columnwidth}
    \includegraphics[width=\columnwidth]{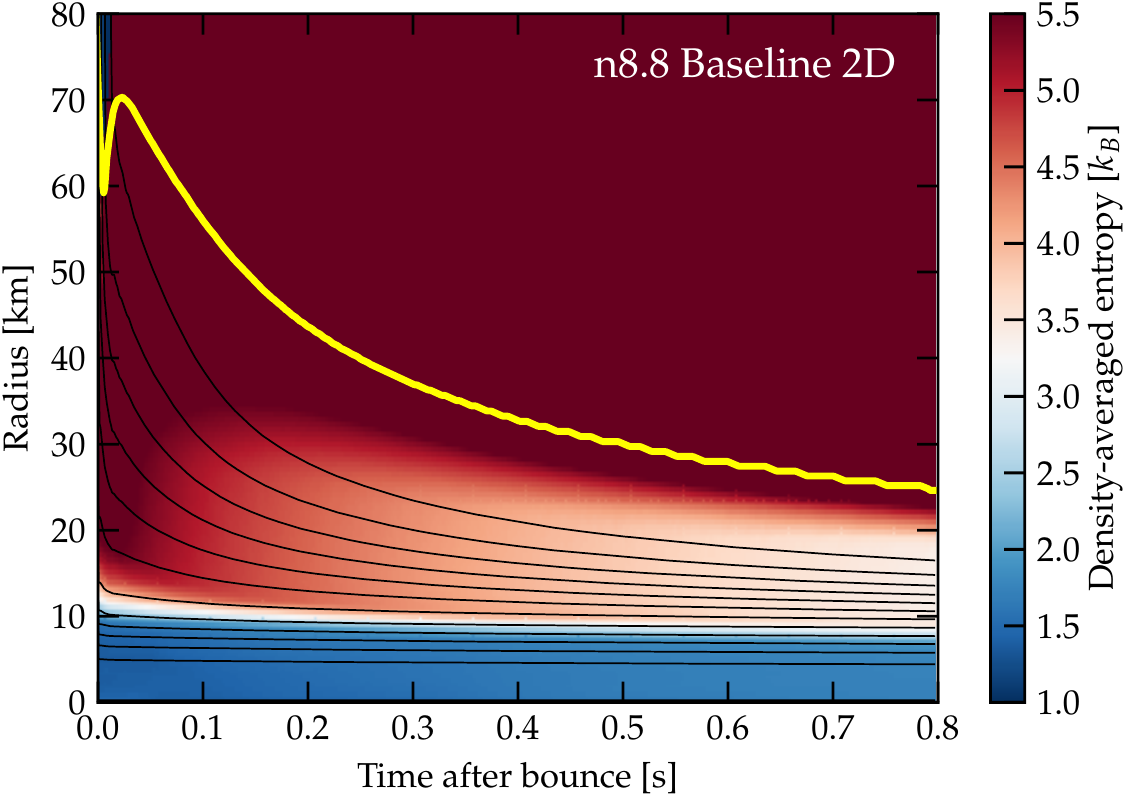}
  \end{minipage}
  \caption{Evolution of the PNS for the n8.8 progenitor in 1D
  (\emph{left panel}) and 2D (\emph{right panel}) with the Baseline
  setup. The black lines are curves of constant enclosed baryonic mass.
  The yellow thick line denotes the PNS radius. The curves are smoothed
  using a running average with a 5-ms window. The background color is
  the density-averaged entropy per baryon in $k_B$. The PNS radius
  contracts to the point of touching the inner convection region,
  visible as the almost constant averaged entropy region exterior to
  ${\sim} 10$ km in 2D, at ${\sim} 0.2\ {\rm s}$ after bounce. As a
  consequence, the subsequent evolution of the PNS is drastically
  different in 1D and 2D.}
  \label{fig:pnssummary}
\end{figure*}

The reason for the faster \ac{PNS} contraction in 1D is more easily
understood considering the n8.8 progenitor with the Baseline setup,
which explodes both in 1D and in 2D. Its \ac{PNS} evolution is shown in
Fig.~\ref{fig:pnssummary}. In the first few hundreds of milliseconds
after bounce, the convection inside the \ac{PNS} is buried deep below
the surface and its impact on supernova evolution is limited, as has
been documented in detail by \citet{buras:2005tb} and
\citet{dessart:2005ck}.  However, over timescales longer than those
considered in either of those works, starting from ${\sim} 0.2\ {\rm
s}$ after bounce, the surface of the \ac{PNS} shrinks to the point of
entering in contact with the inner \ac{PNS} convection, which then
becomes dynamically important.  This can be seen in
Fig.~\ref{fig:pnssummary}, where the region affected by the inner
convection is identifiable by its small radial entropy gradient with
entropy per baryon evolving from ${\sim} 4.5\ {\rm k_B}$ to ${\sim} 3\
{\rm k_B}$ as the \ac{PNS} cools down.

Starting from this moment, the 1D and 2D evolutions begin to diverge. In
1D, the neutrino cooling of the surface is not compensated by convection
and leads to an increasingly steep entropy inversion. The pressure
support in the exterior layers of the \ac{PNS} drops rapidly and leads
to an increased compactness, with respect to the 2D evolution, of the
regions with densities between $10^{11}$ and $10^{13}\ {\rm g}\ {\rm
cm}^{-3}$. These regions are instead inflated in 2D by the deposition of
entropy and lepton number due to convective transport. The structure of
the layers below the inner convection region is also affected, with the
core of the \ac{PNS} reaching higher densities and compactness in 1D.

\begin{figure}
  \includegraphics[width=\columnwidth]{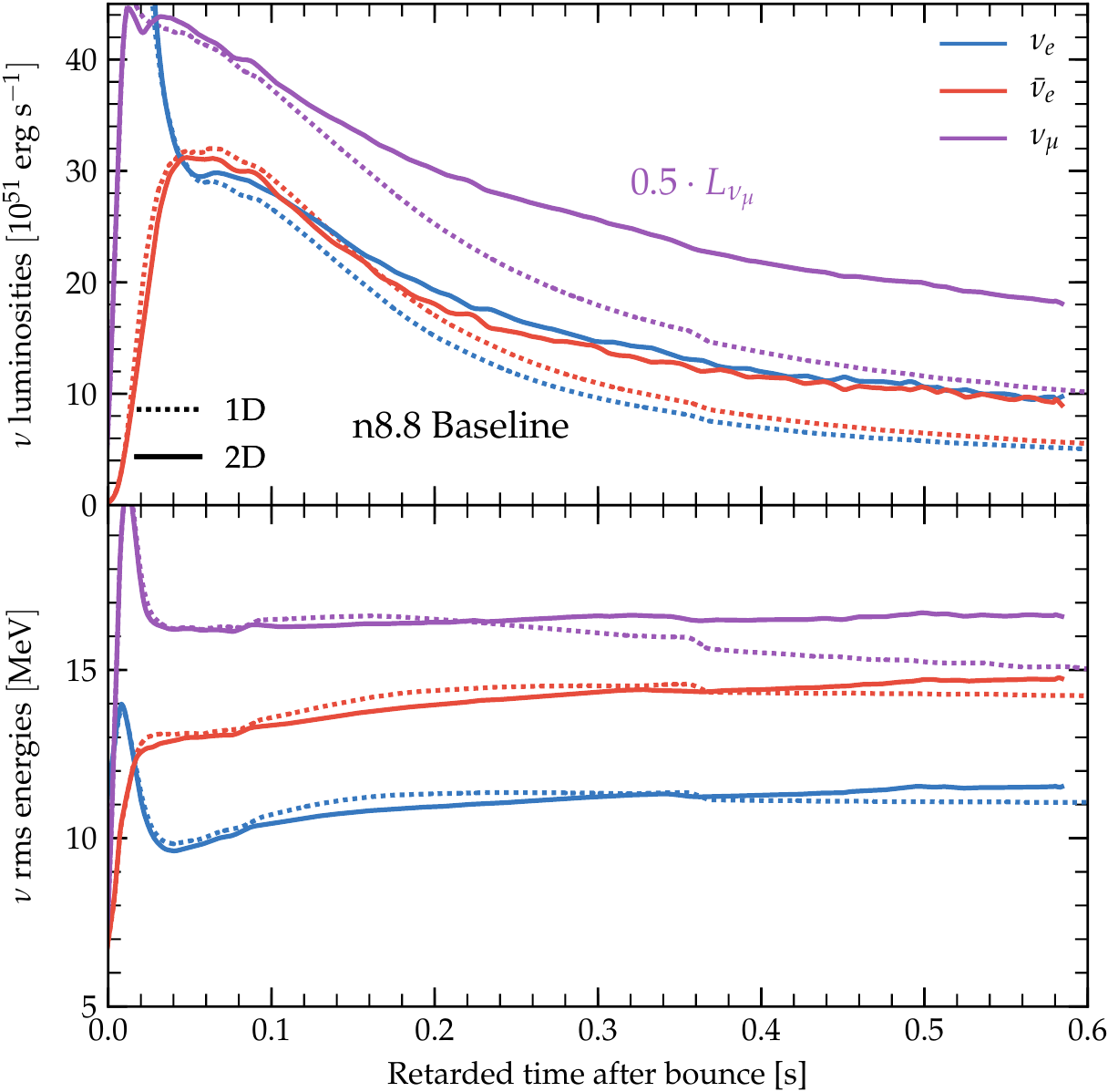}
  \caption{Neutrino luminosity (\emph{top panel}) and rms energies
  (\emph{bottom panel}) at $10,\!000$ km as a function of the retarded
  time for the n8.8 progenitor evolved with the Baseline setup.
  Here, $\nu_\mu$ denotes the sum of all heavy-lepton neutrino species
  and their associated luminosity. It has been rescaled by a factor
  $0.5$ to improve the readability of the plot. The curves are smoothed
  using a running average with a 5-ms window. Neutrino-driven convection
  below the neutrinospheres result in a boost of the luminosity compared
  to 1D models.}
  \label{fig:luminosity.n8p8}
\end{figure}

\ac{PNS} convection also leaves a strong imprint on the neutrino
luminosity, which is boosted by up to a factor ${\sim} 2$ at late times
($\gtrsim 0.5\ {\rm s}$), as can be seen from
Fig.~\ref{fig:luminosity.n8p8}. This seems to be the main reason for the
enhanced growth of the explosion energy for the n8.8, u8.1, and z9.6
progenitors in 2D at late times. While this accounts only for a
relatively small fraction of the explosion energy (see
Fig.~\ref{fig:explene}), the role of \ac{PNS} convection might be more
important for more massive progenitors that explode later in
multi-dimensional simulations.

We remark that \citet{oconnor:2015rwy} also reported modest increases in
the heavy-lepton neutrino luminosities due to \ac{PNS} convection.
However, since they considered models that did not explode in 1D, they
might have underestimated the effect of convection, since the
$\bar{\nu}_e$ and, in particular, the $\nu_e$ luminosities are
significantly affected by accretion. Indeed, while the z9.6 and u8.1
progenitor evolutions (the former only with the ManyBody setup) show
very similar differences between 1D and 2D as does the n8.8, this is not
the case for the 9.0-, 10.0-, and 11.0-$M_\odot$ progenitors, which show
more similar luminosities in 1D and 2D.

The n8.8 model was considered in 1D by \citet{fischer:2009af}, who found
essentially the same luminosity as in our 1D n8.8-Baseline model
($L_{\nu_e}\simeq\ 6\cdot 10^{52}\ {\rm erg}\ {\rm s}^{-1}$ at $0.6\
{\rm s}$ after bounce). This is, however, a factor ${\sim} 2$ smaller
than in our 2D calculations for the n8.8. \citet{mueller:2014rna} also
considered the z9.6 progenitors over a long timescale and found
luminosities very close to ours with the Baseline setup. Their
luminosity was $L_{\nu_e} \sim 10^{52}\ {\rm erg}\ {\rm s}^{-1}$ at
$0.6\ {\rm s}$ after bounce, the same value we also find
(Fig.~\ref{fig:luminosity}), but they did not present a comparison with
the corresponding 1D evolution.  These considerations are all additional
indirect confirmation that the impact of \ac{PNS} convection has been
underestimated.


\section{Conclusions}
\label{sec:conclusions}
We have revisited the explosion of low-mass iron-core \acp{SN} and
O-Ne-Mg gravitational-collapse \acp{SN} with a new set of
neutrino-radiation hydrodynamics simulations in 1D (spherical symmetry)
and in 2D (axial symmetry). Our simulations included the effects of
general relativity in an approximate way and state-of-the-art
multidimensional neutrino transport and weak reactions. Of the six
progenitors we have considered, one, the n8.8 from \citet{nomoto:1984a,
nomoto:1987a}, is the prototype of an \ac{ECSN}.  Two, the $10^{-4}\
Z_\odot$ and zero metallicity u8.1 and z9.6 from Heger (private
communication), have iron cores, but a structure similar to that of the
n8.8. The 9.0-, 10.0-, and 11.0-$M_\odot$ solar-metallicity progenitors from
\citet{sukhbold:2016a} share some similarities with the n8.8, but are
overall closer to the ``canonical'' \ac{CCSN} progenitors considered in
in the \ac{CCSN}-mechanism literature.

As in previous studies \citep{kitaura:2005bt, janka:2007di,
burrows:2007a, fischer:2009af, mueller:2012ak, janka:2012sb,
mueller:2012sv, mueller:2014rna, melson:2015tia, wanajo:2017cyq}, we
find that the n8.8 and z9.6 progenitors typically explode easily, even
in 1D. The u8.1 progenitor is close to the threshold for explosion in 1D
and is successful when many-body corrections to neutral current
reactions are included, as in our ManyBody setup. On the other hand, the
low-mass, but solar-metallicity, progenitors with iron cores from
\citet{sukhbold:2016a} do not explode in 1D and, in some cases, like the
10.0-$M_\odot$, also fail to explode in 2D. Our results show that
solar-metallicity iron-core \acp{SN} do not explode in 1D and are not
even necessarily easier to explode than higher-mass stars.

The failure to explode of the 10.0-$M_\odot$ progenitor from
\citet{sukhbold:2016a} reported here is surprising in the light of the
successful explosion of other progenitors, like the 9.0- and
11.0-$M_\odot$ progenitors from \citet{sukhbold:2016a}. Our findings are
in tension with explodability criteria related to the \ac{ZAMS} mass
\citep{heger:2002by} or to the progenitor compactness and related
parameters \citep{oconnor:2010moj, oconnor:2012bsj, ugliano:2012kq,
nakamura:2014caa, pejcha:2014wda, ertl:2015rga}. Other circumstantial
evidence of limitations in the existing explodability criteria is the
order in which explosions develop in \citet{summa:2015nyk} and the
results of \citet{oconnor:2015rwy}. The latter considered the 12-, 15-,
20-, and 25-$M_\odot$ progenitors from \citet{woosley:2007as} and found
explosions in approximate GR for all progenitors except the $12\
M_\odot$.  Taken together, all of these results suggest that, while some
properties of the explosions are correlated with the compactness of the
progenitor \citep[\eg,][]{oconnor:2012bsj, nakamura:2014caa}, the
explodability is not. Whether an explosion is successful or not depends
on a competition between accretion and neutrino heating
\citep{burrows:1993pi, janka:2000bt, suwa:2014sqa, murphy:2015uwa,
gabay:2015fma} which, in our opinion, has yet to be expressed in terms
of the progenitor properties in a satisfactory way.

We have systematically studied the effect of perturbations and of
changes in the treatment of neutrino-matter interactions, with emphasis
on the impact of the many-body corrections to the neutrino-nucleon
scattering cross section derived by \citet{horowitz:2016gul}. We have
found that relatively small changes are amplified by the proximity to
the threshold for explosion and can lead to qualitatively different
outcomes. For instance, the 10.0-$M_\odot$ model turns from a dud into
an explosion with the inclusion of perturbations in combination with
many-body corrections. This sensitivity to initial conditions and/or
physics setup applies also in 1D to those progenitors that are
sufficiently close to the threshold for explosion. For example, the u8.1
fails to explode in 1D without the inclusion of many-body corrections,
but succeeds when those are included. The reason for the diverging
outcomes with different microphysical suites is easily understood in
terms of the neutrino-radiation intensity and hardness, which directly
translate into the efficiency of the energy deposition by neutrinos.

We have estimated explosion energies by following the development of the
explosions over long timescales and until the shock has reached 19,000
km in 2D simulations. While the explosion energy for the 11.0-$M_\odot$
progenitor is still far from saturated, for the others we have found
saturated explosion energies of the order of a tenth of a Bethe. These
values are in the expected range for low-mass progenitors
\citep{utrobin:2013qna, spiro:2014a, sukhbold:2016a}. We remark that,
while the \ac{ECSN}/\ac{ECSN}-like explosions are nearly-spherical and
we do not expect their explosion energies to change significantly in 3D
\citep{melson:2015tia}, it is likely that the asymmetric explosions we
observe for the 9.0-, 10.0-, and 11.0-$M_\odot$ progenitors will be
quantitatively different in 3D \citep{muller:2015dia}.

For the progenitors that have been evolved in the past by other groups
(the n8.8, u8.1, and z9.6) we obtain explosion energies in 2D that are
typically ${\sim} 50\%$ larger than those of the Garching group
\citep{janka:2007di, melson:2015tia, wanajo:2017cyq}. On the other hand,
we find good agreement with the Garching results for the explosion
energy of the z9.6 progenitor in 1D, which suggests that these
difference might be ascribed to multi-dimensional effects, such as the
handling of convection inside the \ac{PNS} and behind the shock and the
problematic ray-by-ray approximation for lateral neutrino transport
\citep{skinner:2015uhw}.  We also remark that our calculations did not
include nuclear burning and instead assumed nuclear statistical
equilibrium, which could also explain some of the differences. We did
not attempt to systematically investigate the reasons for the
discrepancies, but suggest that the use of models that explode in
self-consistent 1D simulations, in the context of studies on the impact
of the microphysics on the explosion mechanism, or within a renewed
effort to cross-validate \ac{CCSN} codes, appears promising.

For exploding models, we have found final \ac{PNS} masses that have
reached saturation within the simulation time, with the exception of the
11.0-$M_\odot$ progenitor. We find that \acp{ECSN} can explain the
low-mass tail of the observed \ac{NS} mass distribution. The n8.8
progenitor with the ManyBody setup leaves behind a \ac{NS} with baryonic
(gravitational) mass of $1.294\ M_\odot$ ($1.188\ M_\odot$), very close
to the lowest accurately measured \ac{NS} gravitational mass of $1.174
\pm 0.004\ M_\odot$ \citep{martinez:2015mya}.

We studied the evolution of the \ac{PNS}, focusing on those progenitors
that explode both in 1D and 2D. These have nearly identical boundary
conditions, allowing us to quantify the role of multi-dimensional
effects on the long-term evolution of the \ac{PNS}. We have found that
the \ac{PNS} contraction rate slows down significantly in 2D compared to
1D, starting from ${\sim} 0.2\ {\rm s}$ after bounce. At this time, the
\ac{PNS} surface has contracted sufficiently to enter in contact with
the inner \ac{PNS} convection region. The transport of lepton number and
thermal energy by the \ac{PNS} convection then inflates the region  with
densities between $10^{11}$ and $10^{13}\ {\rm g}\ {\rm cm}^{-3}$,
causing a decrease of the contraction rate.

We have also found \ac{PNS} convection to be responsible for a boost of
the neutrino luminosities for all species by up to a factor ${\sim} 2$
at late times $\gtrsim 0.5\ {\rm s}$. This contributed only a $\lesssim
10\%$ increase to the explosion energy for the \ac{ECSN}/\ac{ECSN}-like
progenitor, for which this effect is more easily quantifiable. However,
\ac{PNS} convection is likely to be more important for massive
progenitors that explode late, when the \ac{PNS} surface has already
receded sufficiently close to the \ac{PNS} convection region. Our
results, together with pieces of evidence from \citet{fischer:2009af}
and \citet{mueller:2014rna}, strongly suggest that the impact of
\ac{PNS} convection has been underestimated in the past
\citep{buras:2005tb, dessart:2005ck}. Our findings provide an additional
reason, beside the need to account for continued accretion at late times
\citep{mueller:2014rna}, for the importance of multi-D simulations in
the modeling of the early neutrino signal from cooling \acp{PNS}
\citep[\eg,][]{fischer:2009af, hudepohl:2010a, roberts:2011yw,
roberts:2012zza, nakazato:2012qf, roberts:2016rsf}. At the same time, we
caution the reader that our simulations did not include in-medium
modifications of charged-current reactions, which will also affect the
quantitative properties of the \ac{PNS} neutrino-cooling light curve,
especially after the first second \citep{burrows:1998ek,
martinezpinedo:2012rb, horowitz:2012us, roberts:2012um}.

The main limitation of this work is the assumption of axisymmetry. This
has been necessary given the large computational cost of 3D simulations
with state-of-the-art microphysics, which prevents a systematic study
with multiple progenitors, as is the present one. This is attested by
the scarcity of 3D simulations with full microphysics
\citep{hanke:2013jat, tamborra:2014aua, melson:2015tia, melson:2015spa,
lentz:2015a, summa:2017wxq}. However, moving to 3D will ultimately be
required and will be a goal of our future work.

\section*{Acknowledgments}
The authors acknowledge Chuck Horowitz, Evan O'Connor, and Todd Thompson
for productive conversations concerning, insight into, and help with the
microphysics, and Ernazar Abdikamalov, Sean M.~Couch, Luke F.~Roberts
and Christian D.~Ott for fruitful discussion on the nature of
core-collapse supernovae. Support was provided by the
Max-Planck/Princeton Center (MPPC) for Plasma Physics (NSF PHY-1523261).
DR gratefully acknowledges support from the Schmidt Fellowship.  AB
acknowledges support from the NSF under award number AST-1714267.  JD
acknowledges support from a Laboratory Directed Research and Development
Early Career Research award at LANL. The authors employed computational
resources provided by the TIGRESS high performance computer center at
Princeton University, which is jointly supported by the Princeton
Institute for Computational Science and Engineering (PICSciE) and the
Princeton University Office of Information Technology and by the
National Energy Research Scientific Computing Center (NERSC), which is
supported by the Office of Science of the US Department of Energy (DOE)
under contract DE-AC03-76SF00098. The authors express their gratitude to
Ted Barnes of the DOE Office of Nuclear Physics for facilitating their
use of NERSC. This paper has been assigned a LANL preprint \#
LA-UR-17-20973.

\bibliography{references}

\acrodef{ADM}{Arnowitt-Deser-Misner}
\acrodef{AMR}{adaptive mesh-refinement}
\acrodef{BH}{black hole}
\acrodef{BBH}{binary black-hole}
\acrodef{BHNS}{black-hole neutron-star}
\acrodef{BNS}{binary neutron star}
\acrodef{CCSN}{core-collapse supernova}
\acrodefplural{CCSN}[CCSNe]{core-collapse supernovae}
\acrodef{CMA}{consistent multi-fluid advection}
\acrodef{DG}{discontinuous Galerkin}
\acrodef{HMNS}{hypermassive neutron star}
\acrodef{EM}{electromagnetic}
\acrodef{ECSN}{electron-capture supernova}
\acrodefplural{ECSN}[ECSNe]{electron-capture supernovae}
\acrodef{ET}{Einstein Telescope}
\acrodef{EOB}{effective-one-body}
\acrodef{EOS}{equation of state}
\acrodefplural{EOS}[EOS]{equations of state}
\acrodef{FF}{fitting factor}
\acrodef{GR}{general relativistic}
\acrodef{GRHD}{general-relativistic hydrodynamics}
\acrodef{GW}{gravitational wave}
\acrodef{LIA}{linear interaction analysis}
\acrodef{LES}{large-eddy simulation}
\acrodefplural{LES}[LES]{large-eddy simulations}
\acrodef{NR}{numerical relativity}
\acrodef{NS}{neutron star}
\acrodef{NSE}{nuclear statistical equilibrium}
\acrodef{PNS}{protoneutron star}
\acrodef{SASI}{standing accretion shock instability}
\acrodef{SGRB}{short gamma-ray burst}
\acrodef{SN}{supernova}
\acrodefplural{SN}[SNe]{supernovae}
\acrodef{SNR}{signal-to-noise ratio}
\acrodef{ZAMS}{zero-age main-sequence}

\end{document}